\documentclass[twocolumn,tighten]{aastex631}

\begin{document}

\shortauthors{Luhman \& Alves de Oliveira}
\shorttitle{New Spectral Class of Brown Dwarfs}

\title{A New Spectral Class of Brown Dwarfs at the Bottom of the IMF in IC 348}

\author{K. L. Luhman}
\affiliation{Department of Astronomy and Astrophysics,
The Pennsylvania State University, University Park, PA 16802, USA;
kll207@psu.edu}
\affiliation{Center for Exoplanets and Habitable Worlds, The
Pennsylvania State University, University Park, PA 16802, USA}

\author{C. Alves de Oliveira}
\affiliation{European Space Agency, European Space Astronomy Centre,
Camino Bajo del Castillo s/n, 28692 Villanueva de la Ca\~{n}ada, Madrid, Spain}

\begin{abstract}
In a previous study, we used JWST to identify
three new brown dwarfs in the center of a nearby star-forming cluster, IC~348.
The faintest object had an estimated mass of 3--4~$M_{\rm Jup}$,
making it a contender for the least massive brown dwarf confirmed with 
spectroscopy. Two of the new brown dwarfs also exhibited absorption features
from an unidentified aliphatic hydrocarbon, which were not predicted
by atmospheric models and were not previously detected 
in atmospheres outside of the solar system. We have used JWST to perform a 
deeper survey for brown dwarfs across a larger field in IC~348. 
We have identified 39 brown dwarf candidates in NIRCam images and have
obtained spectra for 15 of them with NIRSpec, nine of which are classified 
as substellar members of the cluster. The faintest new members have
mass estimates of $\sim2$~$M_{\rm Jup}$,
providing a new constraint on the minimum mass of the IMF.
Two new members ($\sim2$ and 10~$M_{\rm Jup}$) exhibit large excess 
emission from circumstellar disks, demonstrating that they
harbor the raw materials for planet formation. Finally, eight of the nine
new brown dwarfs and one known member that is newly observed with NIRSpec 
show the aforementioned hydrocarbon features. Among the total of 11 brown
dwarfs in IC~348 that have hydrocarbon detections, the features are stronger at
fainter magnitudes, indicating that the hydrocarbon is a natural constituent
of the atmospheres of the coolest newborn brown dwarfs.
We propose a new spectral class ``H" that is defined by the presence of the
3.4~\micron\ fundamental band of the hydrocarbon.

\end{abstract}

\section{Introduction}
\label{sec:intro}

The first known brown dwarfs were identified through a search for companions
to nearby stars \citep{nak95,opp95} and a survey for free-floating
members of the Pleiades open cluster \citep{sta94,reb95,reb96,bas96}.
Soon after those discoveries, spectroscopy in nearby star-forming regions 
revealed members that were cool enough to be substellar according to
evolutionary models \citep{hil97,luh97}. Those regions offered the
opportunity to detect brown dwarfs at particularly low masses since substellar
objects are expected to have higher luminosities at younger ages.
The young cluster IC 348 is one of the best sites for a survey for newborn
brown dwarfs because it is nearby ($\sim$300 pc), rich ($\sim$500 members), 
and only moderately obscured by dust from its molecular cloud 
\citep[$A_V<4$,][]{lad95,her98,her08}.
Over the last three decades, we have used optical and infrared (IR) imaging
and spectroscopy to pursue a thorough survey for the members of IC 348
down to the lowest possible masses
\citep{luh98,luh99,luh03,luh05flam,mue07,alv13,luh16,esp17,luh20ic,luh24ic},
uncovering 71 probable brown dwarfs ($>$M6).
Two additional substellar members were found by \citet{all20}.

The James Webb Space Telescope \citep[JWST,][]{gar23} is capable of detecting
members of nearby star-forming regions down to masses of $\sim1$~$M_{\rm Jup}$.
IC 348 is one of the few nearby young clusters that is sufficiently compact
($D\sim0\fdg5$) for imaging with JWST, which has a small field of view. 
During Cycle 1 of the mission, we used NIRCam \citep{rie05,rie23} to obtain 
images of a $6\farcm0\times4\farcm2$ field in the center of the cluster 
\citep[][hereafter L24]{luh24ic}. We identified brown dwarf
candidates in those data and performed spectroscopy with the Near-Infrared
Spectrograph \citep[NIRSpec,][]{jak22} that confirmed three of them as
young late-type objects, one of which is a strong
contender for the least massive free-floating brown dwarf that has
a spectral classification (3--4~$M_{\rm Jup}$). In addition, two of the new
brown dwarfs exhibited absorption bands from an unidentified aliphatic 
hydrocarbon, which were not predicted by atmospheric models 
and were not previously detected in atmospheres outside of the solar system. 
In this Letter, we seek to test scenarios for the origin of that hydrocarbon 
and to better constrain the substellar mass function and its minimum
mass by using JWST to conduct a deeper survey for brown dwarfs across 
a larger field in IC 348.

\section{Imaging}

\subsection{NIRCam Observations}

NIRCam on JWST simultaneously images nearly identical areas of sky
through short and long wavelength channels (0.7--2.4 and 2.4--5~\micron).
The short/long wavelength (SW/LW) channels contain eight/two 2040$\times$2040 
detector arrays with pixel sizes of $0\farcs031/0\farcs063$. The detectors 
cover two $2\farcm2\times2\farcm2$ fields that are separated by $43\arcsec$.
The set of detectors in each of the two fields is described as a module.
In each module, the four detectors for the short wavelength channel
are separated by $5\arcsec$.

\begin{figure}
\epsscale{0.6}
\plotone{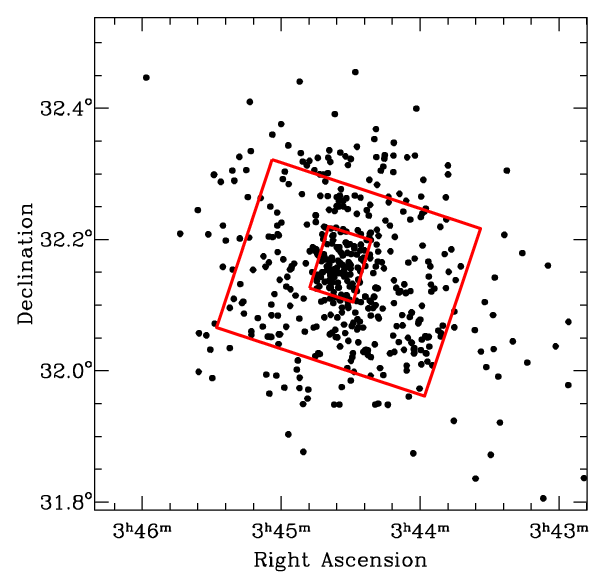}
\caption{
Map of the known members of IC~348 prior to this work and the fields
imaged by JWST/NIRCam in Cycles 1 and 3 (inner and outer rectangles).}
\label{fig:map}
\end{figure}

We obtained NIRCam images of a $16\arcmin\times20\arcmin$ field in IC 348
through program 4866 in August and September of 2024, which was during Cycle 3
of JWST's science operations. The observations were performed with the FULLBOX
dither pattern, four dithers, and a mosaic pattern containing nine rows and 
three columns. Each of the 27 cells in the mosaic was observed in a separate 
visit. In Figure~\ref{fig:map}, we have plotted a map of the members of 
IC 348 that were identified prior to this study (Section~\ref{sec:intro}).
We have marked the boundaries of the fields observed by NIRCam in Cycle 1
(L24) and Cycle 3 (this work). A given position in the new mosaic was
covered by 2--4 exposures per filter. Each exposure utilized five groups, 
one integration, and the SHALLOW4 readout pattern, corresponding to an 
exposure time of 258~s.  We selected the minimum complement of filters 
(two in each channel) for distinguishing brown dwarfs from other astronomical 
sources, which consisted of F162M, F182M, F360M, and F444W. 
The charged time was 41.4 hrs.

\subsection{NIRCam Data Reduction}
\label{sec:nircamred}

To reduce the NIRCam data, we began by retrieving the {\tt uncal} files
from the Mikulski Archive for Space Telescopes (MAST):
\dataset[doi:10.17909/0hke-p124]{http://doi.org/10.17909/0hke-p124}.
We processed the data with version
\dataset[1.16.1]{https://doi.org/10.5281/zenodo.7829329}
of the JWST Science Calibration pipeline under context jwst\_1296.pmap.
We applied detector-level corrections to the {\tt uncal} files
using the {\tt calwebb\_detector1} pipeline module. The resulting count
rate images ({\tt rate}) were processed with the {\tt calwebb\_image2} module,
producing calibrated unrectified exposures ({\tt cal}).

Registration of NIRCam images is performed with the {\tt tweakreg} routine 
within the {\tt calwebb\_image3} module. 
Source-finding algorithms like those within {\tt tweakreg} often produce
spurious detections in areas of extended emission and in the point spread
functions of bright stars, both of which are prevalent in a star-forming
cluster like IC 348.
Therefore, we initially combined the images for a given band into a mosaic with
{\tt tweakreg} deactivated, which assumed that the default world coordinate
systems (WCSs) of the images were correct. We used the {\tt starfind} routine 
in IRAF to identify sources in each mosaic, and we rejected those that
appeared in only one band. The mosaics were visually inspected to remove
remaining spurious detections. The resulting catalog of reliable detections
served as the basis of sources used by {\tt tweakreg} when registering images.

NIRCam images can be registered to an external catalog of sources that
have high precision astrometry, or they can be registered internally using
sources that appear in overlapping images.
High precision astrometry is available from the third data release of the
Gaia mission \citep{gaia16b,bro21,val23}, but the mosaic in IC 348 
encompasses only $\sim100$ non-saturated Gaia sources, so most of the
individual NIRCam images contain few Gaia sources, if any.
Therefore, internal registration of the images was the only option.

When {\tt tweakreg} is applied simultaneously to all of the NIRCam detectors
for a given filter, it solves for a single offset for all of the detectors, 
which means that the relative WCSs for the detectors are assumed to be
accurate. However, noticeable errors are present in the relative WCSs for some
bands \citep{luh24o2}. Ideally, one would avoid such errors by
measuring a separate offset for each detector during registration,
but there is too little overlap among the dithered images, even for the
larger detectors in the LW channel.  The best approach for internal 
registration is to assume that the relative WCSs of the two arrays in 
a LW band are accurate (i.e., measure a single offset for both arrays), 
in which case there is enough overlap among dithered pairs of LW detectors
for internal registration across the entire mosaic. We selected the F360M 
band for this stage of the registration.

We applied {\tt tweakreg} to all {\tt cal} files in F360M, solving for
offsets in both x/y positions and rotation among overlapping images
and solving for global offsets in position and rotation for the entire mosaic
relative to Gaia DR3. We calculated the average
rotation offset for the four dithers in a given visit. Since those dithered 
images should have identical orientations, we used {\tt tweakreg} to rotate 
the {\tt cal} files in a visit by that average offset. We processed
those rotated images with {\tt tweakreg}, solving only for the x/y offsets.
The registered frames in F360M were combined into a mosaic using
the {\tt calwebb\_image3} module.

\begin{figure}
\epsscale{1}
\plotone{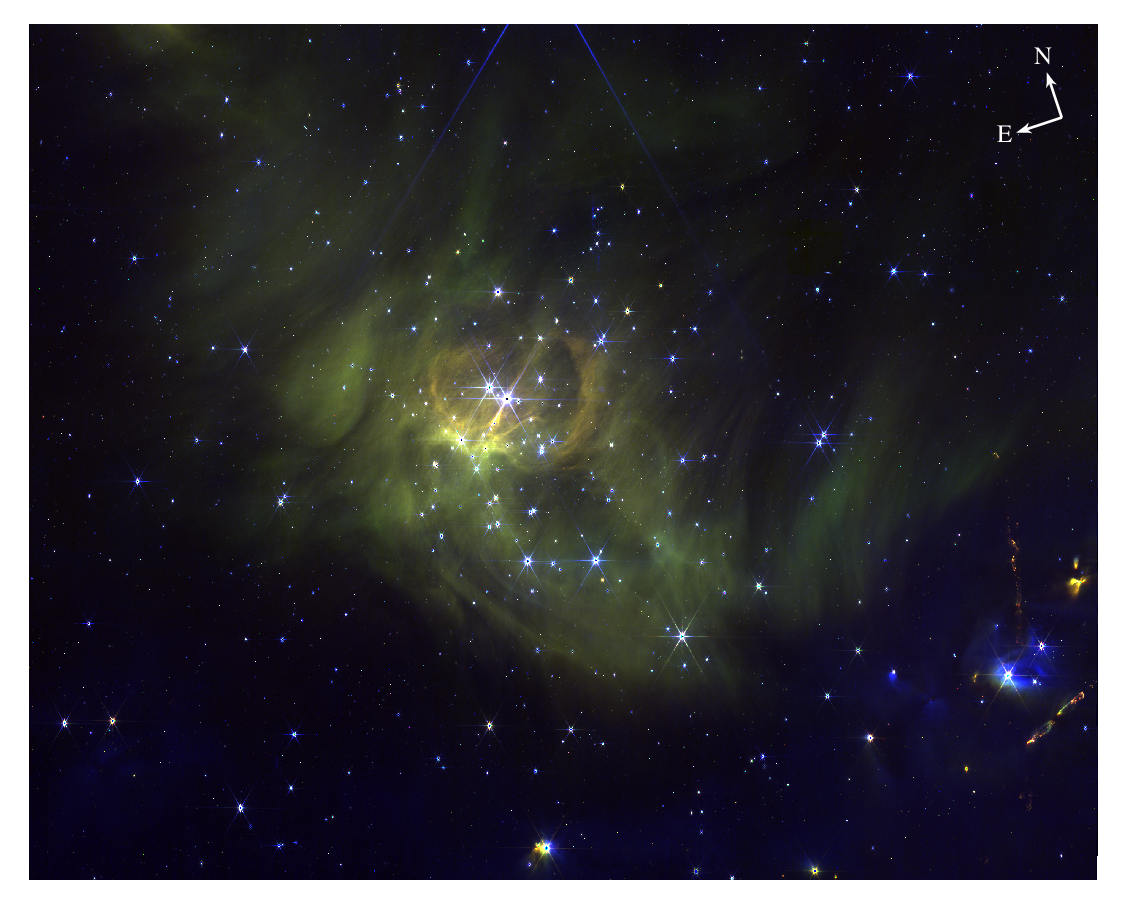}
\caption{
JWST/NIRCam images in three filters (F162M, F360M, F444W) from Cycle 3
for a $16\arcmin\times20\arcmin$ field in IC 348.}
\label{fig:mosaic}
\end{figure}

For {\tt cal} files in F162M, F182M, and F444W, we used {\tt tweakreg} 
to apply the same rotation offsets that were derived for F360M.
For a given visit and module in F444W, we applied {\tt tweakreg} to the 
rotated frames at the four dither positions, solving for relative x/y offsets
and a global x/y offset relative to the source catalog for the final
F360M mosaic. The dithered images for a given detector in the SW bands
do not overlap, so each of those images was aligned individually to the
F360M catalog. The registered images were then combined with 
{\tt calwebb\_image3}. A color composite of the mosaic images for three
of the four filters is presented in Figure~\ref{fig:mosaic}.

We identified sources in the four mosaic images and measured aperture
photometry for them in the manner described by \citet{luh24o2}.
Photometry was measured for aperture radii of 2, 2.5, and 4 pixels.
The band-merged catalog contains $\sim$9400 sources.

\subsection{Identification of Brown Dwarf Candidates}
\label{sec:ident}

\begin{figure}
\epsscale{0.6}
\plotone{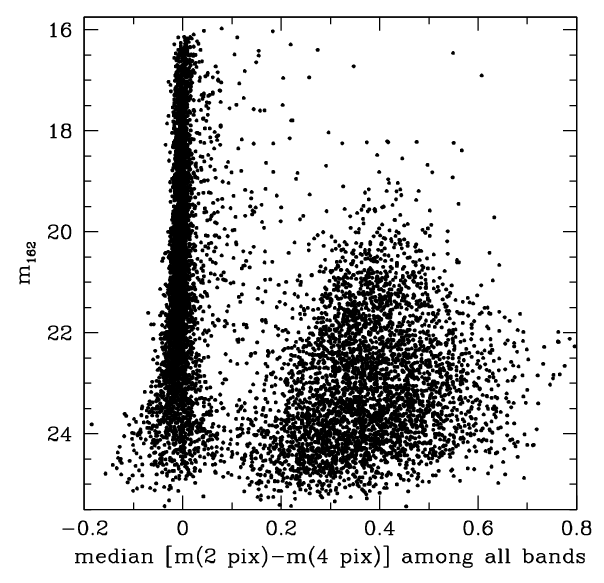}
\caption{
$m_{162}$ versus the median difference between 2 and 4~pixel aperture
photometry among the available bands for sources in NIRCam images of IC~348
from Cycle 3. This metric is used to classify the sources as point-like 
($<0.1$) or extended ($\geq0.1$).
}
\label{fig:cc}
\end{figure}

Many of the sources in the NIRCam images of IC 348 are galaxies.
As done in L24, we have used the differences in photometry between
aperture radii of 2 and 4 pixels to identify sources that are likely to be 
resolved galaxies. In Figure~\ref{fig:cc}, we have plotted $m_{162}$ versus the 
median difference of the 2 and 4 pixel photometry from among the bands in 
which a given source is detected. We classify sources below and above a
value of 0.1 as point-like and extended, respectively.

In Figure~\ref{fig:cmd}, we present color-color and color-magnitude
diagrams in which brown dwarfs can be distinguished from most other
astronomical sources, consisting of $m_{162}-m_{444}$, $m_{162}-m_{182}$,
and $m_{162}$ versus $m_{360}-m_{444}$. We have plotted all point sources 
from the NIRCam images that have nonsaturated detections in all four bands, 
and we have marked the previously known members.
We have selected thresholds in each diagram that capture the known members
with $m_{360}-m_{444}>0.2$ and that should be sufficiently relaxed to
account for reddening, IR excesses, and the undetermined trend of magnitude
versus color for undiscovered members down to the bottom of the mass function. 
The known members with $m_{360}-m_{444}<0.2$, which are missed by our
selection criteria, have spectral types earlier than L0. The previous census 
of IC 348 has a high level of completeness at those spectral types 
\citep[][hereafter L16]{luh16}.

\begin{figure}
\epsscale{0.55}
\plotone{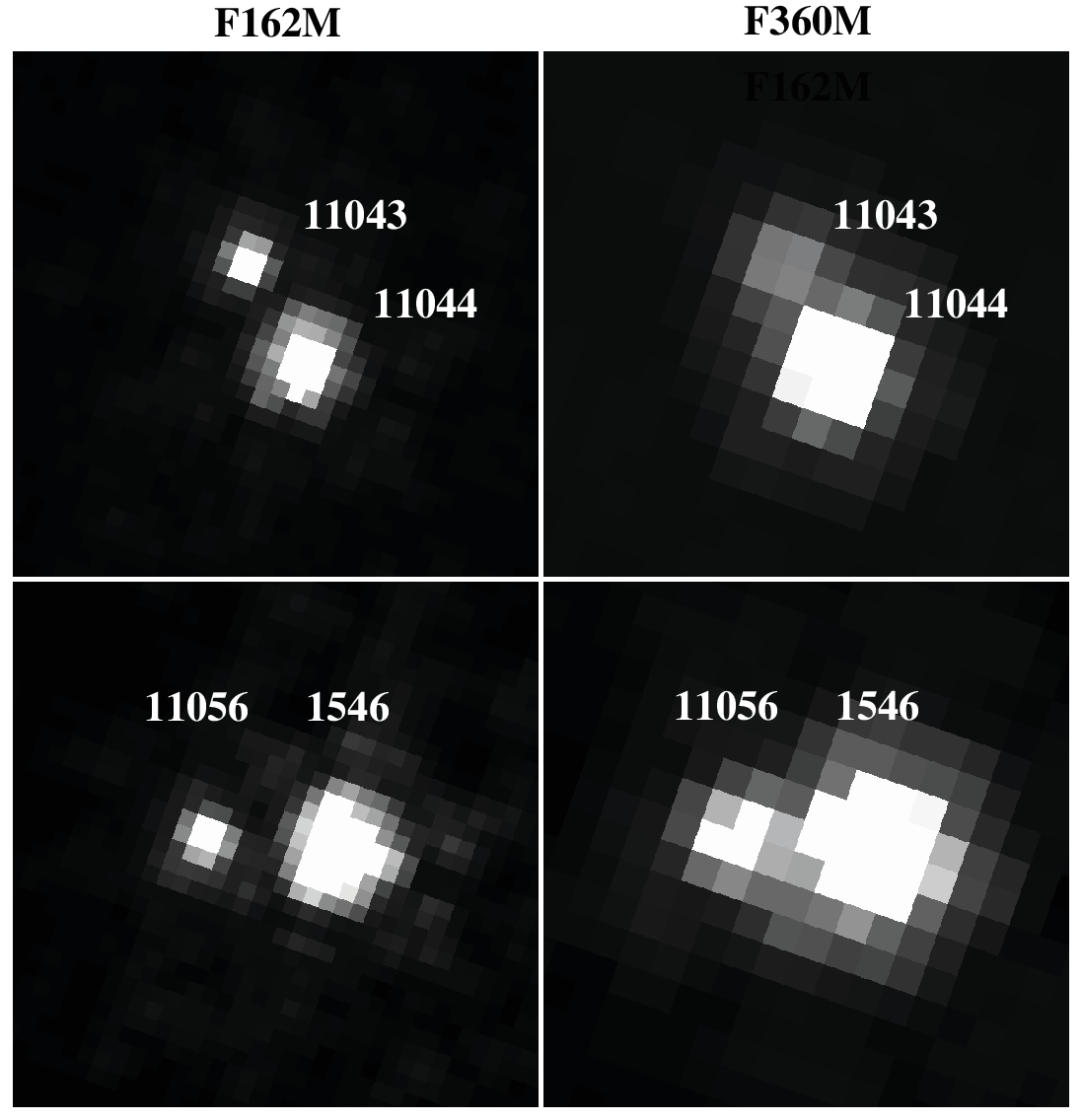}
\caption{JWST/NIRCam images of two close pairs in IC 348.  
LRL 1546 is a previously known brown dwarf while LRL 11043 and LRL 11044 are
classified as new brown dwarfs with NIRSpec (Figure~\ref{fig:spec1}).
LRL 11056 lacks spectroscopy, but its colors are consistent with
a late spectral type. The size of each image is $1\arcsec\times1\arcsec$. 
North is up and east is left.}
\label{fig:bin}
\end{figure}

We have identified the point sources 
that satisfy all of the thresholds in Figure~\ref{fig:cmd}.
A star that is blended with other objects can appear to be extended based 
on the metric in Figure~\ref{fig:cc}, so we have identified sources that are 
flagged as extended and that satisfy our photometric selection criteria, 
and we have inspected their images to check if they are blended point
sources. In addition, we have inspected the images of all candidates
selected from Figure~\ref{fig:cmd} to verify that they are point sources
and check for close companions or contamination from bad pixels.
One of the candidates (LRL 11044) does have a possible secondary companion at
a separation of $0\farcs23$ (LRL 11043). 
We also inspected the images of previously known members that
are not saturated, resulting in the identification of a candidate companion
at a separation of $0\farcs27$ from LRL 1546 (L0). The two pairs are 
well-resolved in the SW images and are partially blended in the LW images, 
as shown in Figure~\ref{fig:bin}. In each of the LW bands, we fit and 
subtracted the point spread function of a given component to more accurately 
measure photometry for its companion. Using the resulting photometry, all
components of the two pairs satisfy our selection criteria for brown dwarfs.

Our final sample of brown dwarf candidates contains 39 sources.
In Figure~\ref{fig:cmd}, we have marked those candidates with symbols
that are based on their spectral classifications from NIRSpec 
(Section~\ref{sec:spec}) or the absence of NIRSpec data.
In Table~\ref{tab:phot}, we present astrometry and photometry for the 39
candidates and the 43 previously known members that have a nonsaturated
detection in at least one band.

\subsection{Proper Motions}
\label{sec:pm}

\begin{figure}
\epsscale{0.6}
\plotone{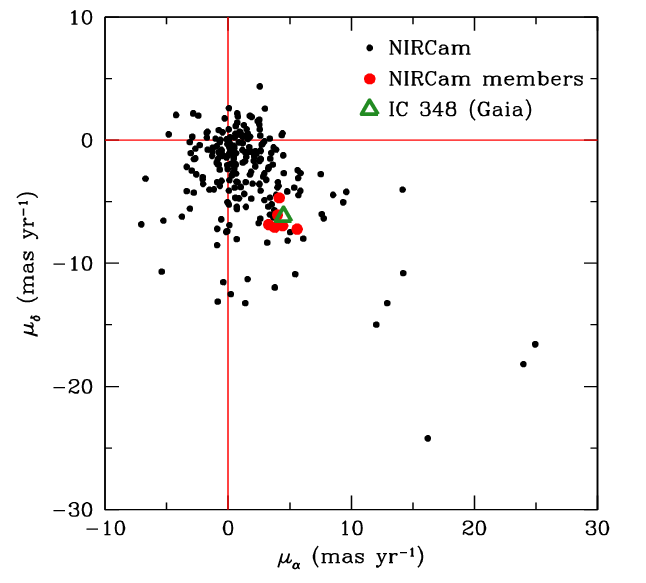}
\caption{
Proper motions measured for point sources in JWST/NIRCam images of IC 348 
from Cycles 1 and 3. We have marked the six previously known members of the
cluster that appear in both epochs and are not saturated, three of which were
discovered with the Cycle 1 data. We also indicate the proper motion of 
the cluster based on Gaia DR3.}
\label{fig:pm}
\end{figure}

Since two epochs of NIRCam images are available for the center of IC 348,
we can use those data to measure proper motions, which provide constraints
on cluster membership. We have reprocessed the images from Cycle 1 with
procedures similar to those applied to the Cycle 3 data in
Section~\ref{sec:nircamred}. For the Cycle 1 images in F360M, we used
the catalog of galaxies from the F360M mosaic in Cycle 3 for the
absolute calibration of the astrometry, which enables the measurement of
absolute proper motions. The pair of dithered images
for a given module was aligned to the Cycle 3 mosaic using 30--60 galaxies.
We then aligned the other Cycle 1 bands to 
the F360M mosaic in Cycle 1 using both stars and galaxies.
In Figure~\ref{fig:pm}, we have plotted the proper motions measured for
point sources from the two epochs of NIRCam data. 
Based on kinematic models of the Milky Way \citep{rob03}, most field stars in
the direction of IC 348 and in the magnitude range of the NIRCam data
should have proper motions of 1--5 mas~yr$^{-1}$, which is consistent
with the measurements in Figure~\ref{fig:pm}. We have marked the
six previously adopted members of the cluster that appear in both epochs and
that are not saturated, which consist of LRL 595, LRL 596, LRL 621, and
the three objects found in L24 (LRL 11001, LRL 11003, LRL 11004). 
In addition, we have indicated in Figure~\ref{fig:pm} the median
proper motion from Gaia DR3 for the 40 stars in IC 348 that have
the most accurate parallaxes, which corresponds to
$(\mu_{\alpha}, \mu_{\delta}=4.52,-6.23)$~mas~yr$^{-1}$.
The standard deviation of each component is $\sim0.6$~mas~yr$^{-1}$.
The six previous members have proper motions 
that are tightly grouped near the Gaia measurement for IC 348, which supports
their membership in the cluster. 
The median NIRCam motion for the six members is
$(\mu_{\alpha}, \mu_{\delta}=4.1,-6.9)$~mas~yr$^{-1}$ and the standard
deviation in each component is  $\sim0.9$~mas~yr$^{-1}$.
None of our new brown dwarf candidates
from Cycle 3 appear within the field imaged in Cycle 1.

\section{Spectroscopy}
\label{sec:spec}

\subsection{NIRSpec Observations}

We observed a subset of the brown dwarf candidates
from Section~\ref{sec:ident} using the multi-object spectroscopy mode 
of NIRSpec on JWST \citep{fer22} through program 4866 in February of 2025.
That mode employs the microshutter assembly (MSA), which consists of
four quadrants that each contain 365$\times$171 shutters.
An individual shutter has a size of $0\farcs2\times0\farcs46$.
The MSA covers a field with a size of $3\farcm6\times3\farcm4$.
Spectra are collected with two $2048\times2048$ detector arrays that have
pixel sizes of $0\farcs103\times0\farcs105$. We selected the PRISM disperser
for our observations because of its sensitivity and wide wavelength coverage
(0.6--5.3~\micron).  The spectral resolution for PRISM data ranges 
from $\sim$40 to 300 from shorter to longer wavelengths.  

The NIRSpec observations were divided among four visits, each of which
was assigned an aperture position angle by the observatory.  We manually
searched for pointings in which the detector arrays would encompass large 
numbers of candidates for the four position angles.
For each of those fields, we used the MSA Planning Tool (MPT)
in the Astronomer's Proposal Tool (APT) to search a small area 
($\sim20\arcsec\times20\arcsec$) for a pointing that would optimize
the sample of observable targets, giving higher priority to fainter
candidates. In addition to the brown dwarf candidates, we included 
``filler" targets at low priority, which consisted of known late-type
members of IC 348, candidates for highly reddened background stars,
and sources with $m_{360}-m_{444}>0.45$ that did not satisfy the
criteria for brown dwarf candidates (Figure~\ref{fig:cmd}).

When designing the MSA configurations with MPT, we required that targets
would be observed at three nod positions in three adjacent shutters, 
which are equivalent to a single $0\farcs2\times1\farcs5$ slitlet. 
For one of the selected pointings, LRL 22705 (a known member) was 
observable in only two shutters while the third shutter was stuck closed,
so it was not automatically selected as a target by MPT. We manually opened
the two operable shutters for that object.  As mentioned in 
Section~\ref{sec:ident}, two of the brown dwarf candidates
(LRL 11043 and LRL 11044), form a close pair ($0\farcs23$), and another 
candidate (LRL 11056) has a separation
of $0\farcs27$ from a known substellar member (LRL 1546).
The components of the $0\farcs23$ pair were observed in one
of our four MSA configurations, aligned along the slitlet 
and straddling the bar between adjacent shutters.
We manually opened an extra shutter so that each component would appear
in an open shutter for all three nod positions.  The primary for the 
$0\farcs27$ pair, LRL 1546, was one of our filler targets. The position angle
of its secondary placed it outside of the slitlet during that observation.

We obtained usable spectra for 15 brown dwarf candidates and eight
filler targets. Those sources are listed in Table~\ref{tab:spec}.
One additional target, LRL 11026, was contaminated by light from a 
nearby bright star, so we were unable to measure a reliable spectrum.

For each of the four MSA configurations, we collected one 
exposure at each of the three nod positions. For two configurations,
a single exposure utilized 100 groups, one integration, and the
NRSIRS2RAPID readout pattern. To avoid saturation for the brightest
targets in the remaining two configurations, we used 68/45 groups and 
two/three integrations, both with the NRSRAPID readout pattern.
The total exposure time for each configuration was $\sim$4400 s.
The charged time was 10.3 hrs.

\subsection{NIRSpec Data Reduction}

\begin{figure}
\epsscale{0.6}
\plotone{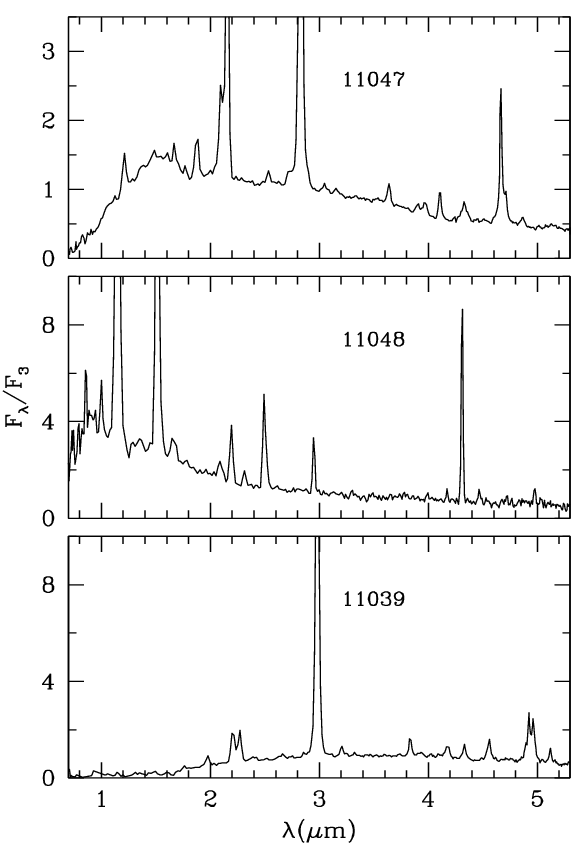}
\caption{JWST/NIRSpec spectra for two brown dwarf candidates in IC 348
and a filler target that are active galaxies.}
\label{fig:spec3}
\end{figure}

The data reduction for NIRSpec began with the retrieval of the
{\tt uncal} files from MAST:
\dataset[doi:10.17909/xxg7-hc43]{http://doi.org/10.17909/xxg7-hc43}.
As done in L24, we performed two separate reductions of the data
that used the JWST Science Calibration pipeline and 
a custom version of the pipeline developed by the ESA NIRSpec
Science Operations Team \citep{alv18,fer22}.
In L24, we preferred the reduced spectra from the latter pipeline.
However, we find that the latest version of the JWST Science Calibration
pipeline (\dataset[1.17.1]{https://doi.org/10.5281/zenodo.7829329})
produces spectra that are comparable to those from the ESA NIRSpec pipeline.
We have adopted the results from the JWST Science Calibration pipeline.
The reduced spectra are presented in Figures~\ref{fig:spec1}--\ref{fig:spec3}.
We have used the same pipeline to reprocess the NIRSpec data for the three
brown dwarfs found in L24. Those updated reductions are included in
Figure~\ref{fig:spec1}. All spectra have been flux calibrated 
using the photometry from NIRCam.

\subsection{Spectral Classifications} 
\label{sec:class}

\subsubsection{Nonmembers}

\begin{figure}
\epsscale{1.2}
\plotone{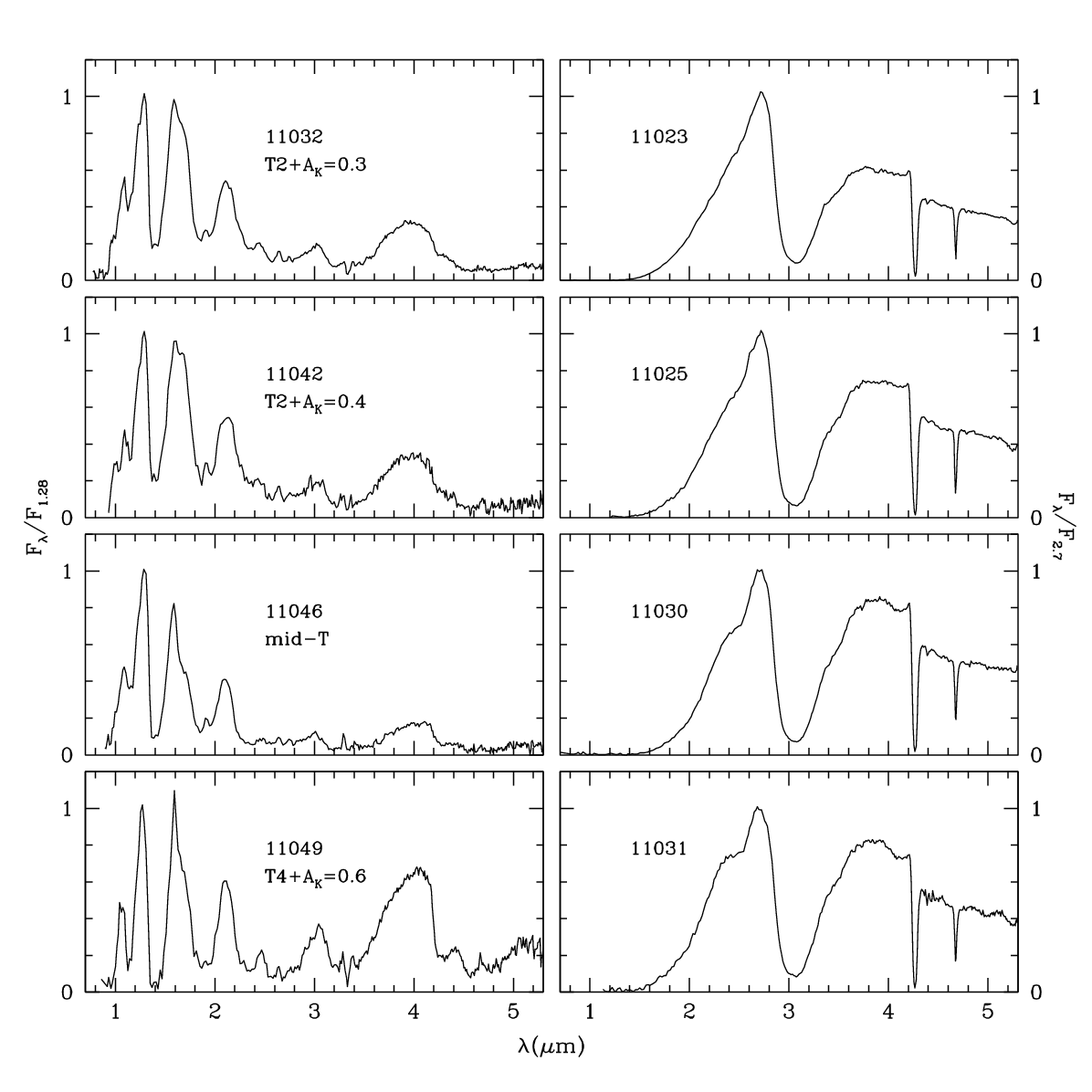}
\caption{JWST/NIRSpec spectra for brown dwarf candidates in IC 348 that are
background T dwarfs (left) and highly reddened background stars that were
selected as filler targets (right).}
\label{fig:spec2}
\end{figure}

Four candidates are T dwarfs based on the
presence of absorption bands from methane (Figure~\ref{fig:spec2}).
The fundamental band of CO at 4.4--5.2~\micron\ is much weaker in the youngest
L dwarfs than in old L dwarfs in the field \citep{luh23}, so the strong 
CO bands in the T dwarfs observed by NIRSpec suggest that they are field 
dwarfs rather than members of IC 348. We have measured spectral types for the 
T dwarfs through comparison to standard spectra at 1--2.5~\micron\ for a range 
of reddenings \citep{bur06}. One of the objects, LRL 11046, is not matched
well by any reddened standard spectrum, but its absorption bands are
suggestive of mid-T types. That object is both the reddest in 
$m_{360}-m_{444}$ and bluest in $m_{162}-m_{444}$ among the four T dwarfs 
in the NIRSpec sample.  We estimated distances of 500--800 pc for the T dwarfs 
based on their photometry and the typical absolute magnitudes for T dwarfs 
at their spectral types \citep{dup12}. The four candidates for reddened 
background stars that were selected as filler targets do show red spectral 
slopes and deep ice features that indicate high extinctions 
(Figure~\ref{fig:spec2}). They are not protostars since
they lack IR excesses in photometry from NIRCam and the Spitzer Space
Telescope \citep{lad06,mue07}. Two brown dwarf candidates and one filler 
target are active galaxies based on their redshifted emission lines 
(Figure~\ref{fig:spec3}). 

\subsubsection{New Members}

The nine remaining brown dwarf candidates observed by NIRSpec have
late spectral types based on their H$_2$O absorption bands.
The triangular $H$-band continua and weak CO bands in these candidates
indicate that they are young \citep{luc01,luh23}, and therefore are 
likely members of IC 348. As a reminder, the three brown dwarfs
found in L24 were classified as young members based on the same age
diagnostics, and our measurements of their proper motions have 
further established their membership (Section~\ref{sec:pm}).
The NIRspec data for two candidates (LRL 2296, LRL 11040) contain large
excess emission at $>$3.5~\micron\ from circumstellar disks, providing
additional evidence of youth.

The spectral types of the new members are discussed in Section~\ref{sec:newclass}.
Table~\ref{tab:spec} contains the membership status and spectral
classification for each NIRSpec target. It also includes redshifts
measured from the galaxy emission lines (Donald Schneider, private
communication).

\subsection{Candidates that Lack Spectroscopy}
\label{sec:viable}

The NIRSpec observations provide some indication of the viability of the
remaining brown dwarf candidates that lack spectroscopy.
Seven of the original candidates have negative colors in $m_{162}-m_{182}$
that are indicative of T dwarfs. Since four of them were found to be
background T dwarfs with NIRSpec, it seems likely that the same is true
for the other three candidates (LRL 11063, LRL 11066, LRL 11071). 
LRL 11026 and LRL 11055 resemble the galaxies in the NIRSpec sample in
their colors, and the former may be slightly extended.
The remaining candidates more closely match the confirmed members,
although a few are somewhat redder in $m_{162}-m_{182}$,
which could reflect higher reddenings. Two of the viable candidates that lack 
spectra (LRL 11064, LRL 11072) are fainter than the confirmed members.

\section{A New Spectral Class}
\label{sec:newclass}

\subsection{Origin of the 3.4~\micron\ Feature}

In L24, we found two new members of IC 348 (LRL 11001, LRL 11003) that
exhibited a strong absorption band near 3.4--3.5~\micron, which coincides
with the so-called 3.4~\micron\ feature that has been observed in the diffuse 
interstellar medium (ISM) \citep{soi76}, meteorites and solar system dust
\citep{wdo88,mat05}, and the atmospheres of Saturn and Titan \citep{bel09}.
That feature has been attributed to fundamental stretching modes in an
unidentified aliphatic hydrocarbon\footnote{We refer to the carrier
in the singular, but multiple species may be responsible for the
3.4~\micron\ feature.} \citep{san91,pen94,pen02}.
In L24, we removed the steam absorption bands of the hydrocarbon-bearing
objects through division by the spectrum of a normal young L dwarf, which
revealed the overtone and combination bands from the hydrocarbon
at near-IR wavelengths. We concluded in L24 that the
hydrocarbon resides in the atmospheres of the brown dwarfs rather than
a foreground medium for the following reasons:
(1) the extinctions of the brown dwarfs are much too low for 
3.4~\micron\ absorption from the diffuse ISM;
(2) the 3.4~\micron\ feature has not been previously detected in molecular
clouds, protostellar envelopes, or edge-on disks; and
(3) the detections of the overtone and combination bands indicate that that 
hydrocarbon resides in atmospheres rather than a foreground medium.

Since model atmospheres for brown dwarfs do not
predict the presence of a non-methane hydrocarbon, we speculated in L24 about
possible scenarios for its origin, such as UV-induced photochemistry.
With our new NIRSpec observations, there are now a total of 11 members
of IC 348 that show significant detections of the 3.4~\micron\ feature 
(Section~\ref{sec:sequence}, Figure~\ref{fig:spec1}). One additional
source, LRL 11004, may also have a marginal detection. Among these objects, the
strength of the feature is correlated with apparent magnitude,
as illustrated in Figure~\ref{fig:ew}. As discussed in 
Section~\ref{sec:sequence}, the NIRSpec data in IC 348 exhibit a continuum of
spectral variations from (1) normal young L dwarfs to
(2) weak 3.4~\micron\ absorption in otherwise normal L-type spectra
to (3) moderate 3.4~\micron\ absorption and changes in the near-IR slopes
and band strengths to (4) strong 3.4~\micron\ absorption and a reversion to
L-type near-IR slopes and band strengths. The correlation between
the 3.4~\micron\ band and apparent magnitude and the presence of a
clear spectral sequence between L dwarfs and objects with that feature
indicate that the hydrocarbon is a natural constituent of the coolest 
brown dwarfs in star-forming regions.

\subsection{A Spectral Sequence for Hydrocarbon-bearing Objects}
\label{sec:sequence}

\begin{figure}
\epsscale{1.2}
\plotone{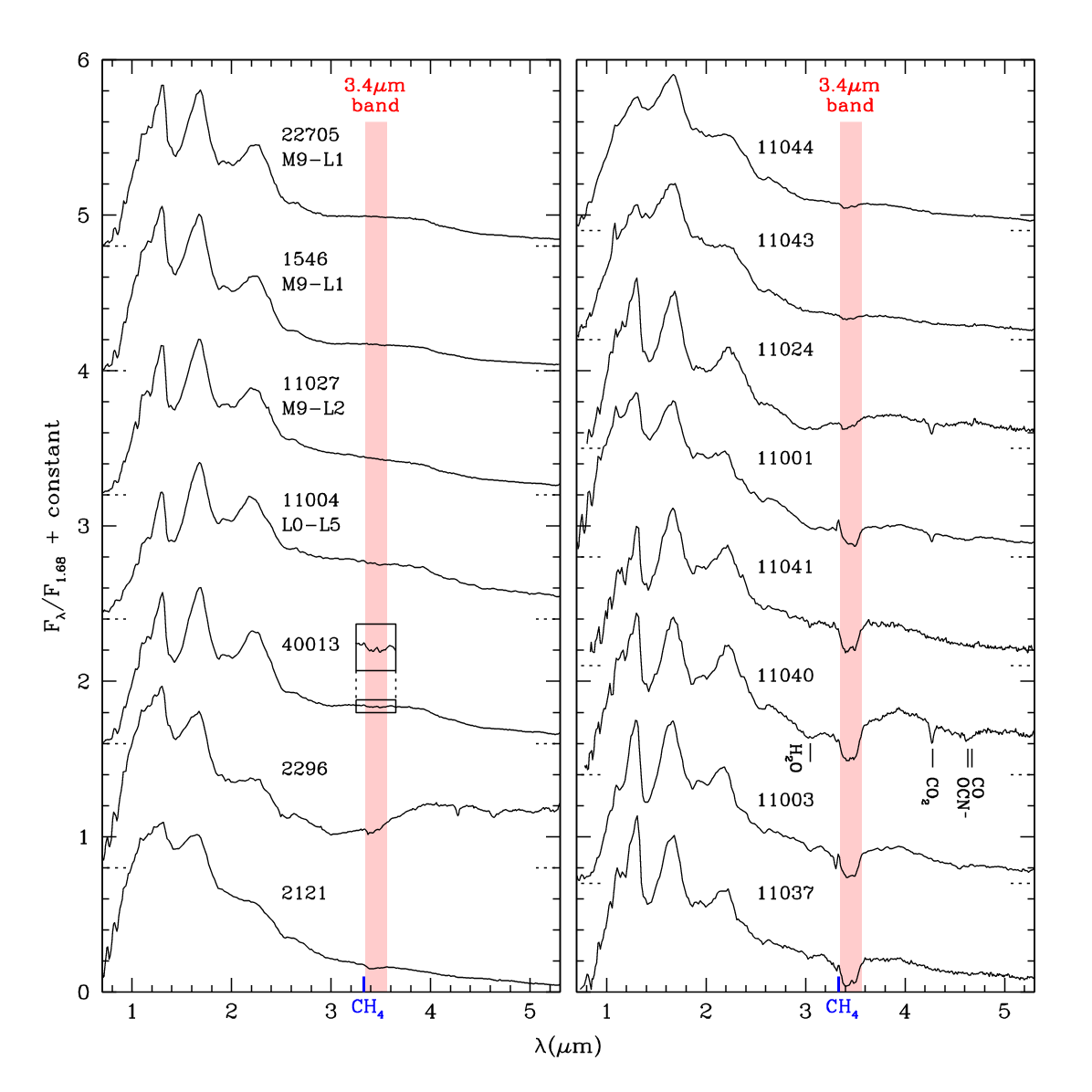}
\caption{JWST/NIRSpec spectra collected in this work and in L24
for brown dwarfs in IC 348. The first four objects are normal young L dwarfs.
The remaining sources exhibit the 3.4~\micron\ feature from an unidentified
aliphatic hydrocarbon. The wavelength range of that feature is marked by 
the pink band. For reference, the central wavelength of 
the Q branch of CH$_4$ (not detected in these objects) is indicated
by the blue tick mark. Two brown dwarfs (LRL 2296, LRL 11040) have large 
excess emission at $>$3.5~\micron\ and absorption features that may arise
from ices (H$_2$O, CO$_2$, OCN$^-$, CO).
For each spectrum that has been shifted upward, the
level for zero flux is marked by a pair of dotted tick marks.}
\label{fig:spec1}
\end{figure}

\begin{figure}
\epsscale{0.6}
\plotone{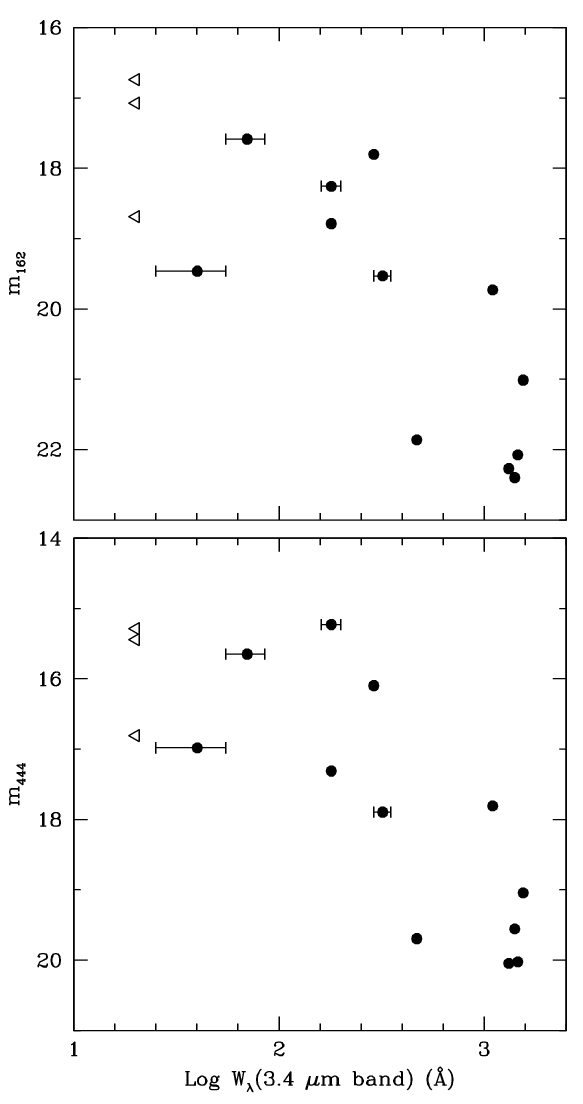}
\caption{Photometry in F162M and F444W versus the equivalent width 
of the 3.4~\micron\ feature for brown dwarfs in IC 348 (Figure~\ref{fig:spec1}).
Error bars are omitted when they are smaller than the symbols.}
\label{fig:ew}
\end{figure}

In Figure~\ref{fig:spec1}, we have presented the NIRSpec spectra 
of members of IC 348 from L24 and this work in order of the strength
of the 3.4~\micron\ feature. The first four objects lack that feature, and
have the spectra of normal young L dwarfs. We have indicated in
Figure~\ref{fig:spec1} the spectral types produced by a comparison
to young standard spectra at 1--2.5~\micron\ \citep{luh17}. 
Because of a degeneracy between spectral type and reddening for
young L dwarfs, the reddest object, LRL 11004, can be matched by a wide
range of standards (e.g., reddened L0, unreddened L5). Although the resolution 
of the NIRSpec data is very low at optical wavelengths, TiO and VO bands 
appear to be detected at 0.7--0.8~\micron\ in the first two objects near L0, 
while they seem to be absent for the next two L dwarfs that are slightly 
later, which is consistent with previous optical spectra of young L dwarfs 
\citep{cru09}.

The 11 objects in Figure~\ref{fig:spec1} that have detections of the 
3.4~\micron\ band exhibit the following broad trends (with a few exceptions):
(1) when the 3.4~\micron\ feature first appears (LRL 40013),
TiO and VO may begin to reemerge while the H$_2$O bands and near-IR spectral
slope remain unchanged relative to normal L dwarfs;
(2) as the 3.4~\micron\ absorption increases, TiO and VO continue
to strengthen, the H$_2$O bands weaken, and the near-IR slopes become bluer;
(3) when the 3.4~\micron\ band is strongest, the H$_2$O bands return to the 
depths of L dwarfs and the near-IR slopes may become somewhat redder.
A comparison of the spectral slopes should account for the possibility
of reddening (i.e., the intrinsic spectra may be bluer than the observed
spectra). Regardless of reddening, it is clear that sources with intermediate 
3.4~\micron\ strengths tend to have the bluest near-IR slopes.
The reversal in spectral features for those objects is reminiscent of a less 
extreme reversal in near-IR colors and FeH strengths across the L/T transition 
among field dwarfs \citep{bur02b}.

The expectation has been that the youngest brown dwarfs would show methane
absorption below some temperature threshold, assuming that the initial
mass function (IMF) extends
to low enough masses. Instead, a different aliphatic hydrocarbon appears
in the coolest newborn brown dwarfs. The challenge is for theoretical
models \citep[e.g.,][]{zha25} to explain the presence of that hydrocarbon, 
the absence of methane, and the other spectral trends that we have observed.

\subsection{The Proposed ``H" Spectral Class}

From earlier to later types, the beginnings of the L/T/Y spectral
classes are defined primarily by the significant weakening of TiO at
red optical wavelengths (L) and the onset of methane (T) and ammonia (Y)
at near-IR wavelengths \citep{mar97,kir99,bur02a,geb02,kir05,cus11}.
Since we have identified a new population of young brown dwarfs that do
not fall within any of these spectral classes, we propose the definition
of a new class that is based solely on the presence of the
3.4~\micron\ band. When considering letters to select for new classes, 
\citet{kir99} suggested that H, L, T, and Y were the best options at that time.
We propose to adopt H for the new class that is tied to the
3.4~\micron\ feature. The letters used for spectral types often do
not carry meaning, but ``H" could stand for ``hydrocarbon" (a non-methane
aliphatic variety). We refrain from attempting to define subclasses and
identify specific objects to serve as standards until a larger sample
of these hydrocarbon-bearing objects is available. 
Depending on the trends among spectral features that
emerge in a larger sample, it may be possible to define H subclasses
through a combination of the 3.4~\micron\ band and features from other 
species like TiO, VO, and H$_2$O.

\section{Properties of the Substellar Population}
\label{sec:imf}

\subsection{Initial Mass Function}

\subsubsection{IMF Sample}

We wish to incorporate the results of our survey for brown dwarfs in IC 348
into a new estimate of the cluster's IMF.
We have combined the compilation of members from L16
with additional members found in subsequent studies
\citep[][L24]{esp17,luh20ic,all20,lal22} and in this work.
We have omitted LRL 322, LRL 390, and WBIS 03441864+3218204,
which we have classified as nonmembers based on astrometry
from Gaia DR3. Only sources with spectral classifications are
considered, which excludes some close companions and a few protostars.
For a population like IC 348 that is associated with a molecular cloud,
the members have a range of extinctions. Since more massive members 
can be detected through greater extinction,
a sample that includes all known members is susceptible to a bias
against members at lower masses. To obtain a sample of members that is
unbiased in mass and that is an accurate reflection of the IMF in the
cluster, it is necessary to define an extinction-limited sample in which 
the extinction
limit is high enough to encompass a large number of members but low
enough that the completeness limit reaches low masses.
In L16, we found that our census of IC 348 within a 
large area that extended beyond the NIRCam field had a high level
of completeness down to masses of $\sim$0.01~$M_\odot$ for extinctions
of $A_J<1.5$ ($A_K<0.57$).
Therefore, we include in our IMF sample the known members that have
$A_J<1.5$ and that are located within the NIRCam field.
The resulting sample contains 281 objects.
We assume that all of the new members found with JWST in L24 and this work
are within the extinction limit for the IMF sample (Section~\ref{sec:histo}).

\subsubsection{Histograms of Absolute Magnitudes}
\label{sec:histo}

\begin{figure}
\epsscale{1.2}
\plotone{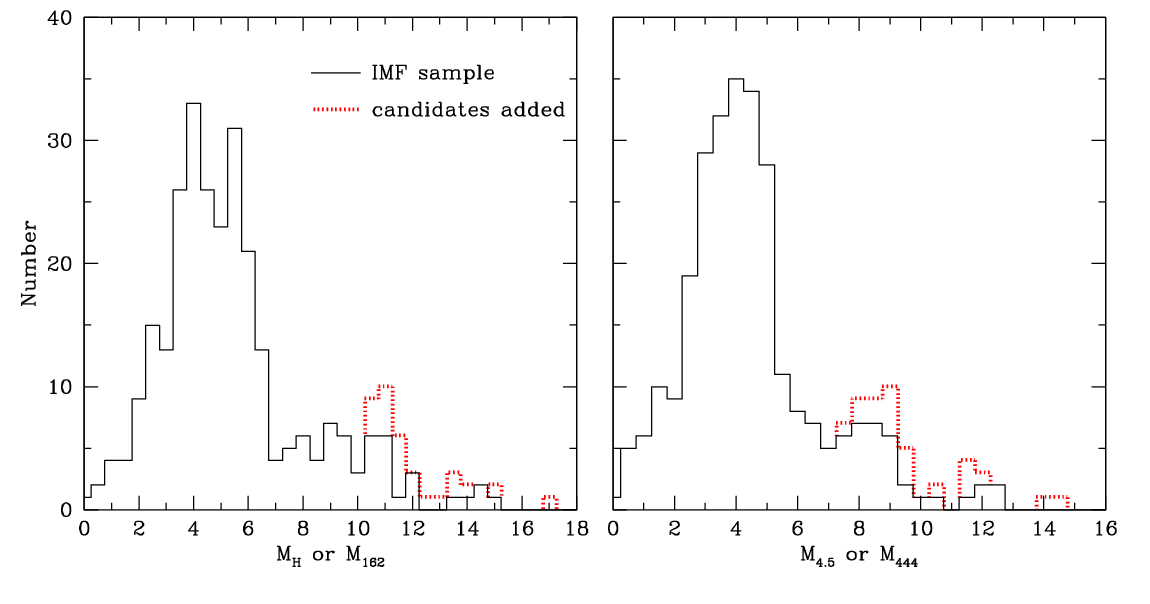}
\caption{Histograms of extinction-corrected absolute magnitudes in 
$H$ or F162M (left) and [4.5] or F444W (right) for known members of IC 348 in 
our IMF sample (solid line) and the remaining viable NIRCam candidates that 
lack spectroscopy (dotted line).}
\label{fig:histo}
\end{figure}

To characterize the IMF in terms of observational parameters
that should be correlated with mass, we have plotted two histograms of
extinction-corrected absolute magnitudes in Figure~\ref{fig:histo} for
two pairs of similar filters.
We have adopted a distance of 313 pc for all members (L24).
Extinctions are taken from L16 when available in that study
and otherwise are derived in the same manner. We normally estimate extinctions 
from a comparison of an observed spectrum or color to the typical intrinsic
color or spectrum for a young photosphere with the spectral type in question,
but those typical properties are undetermined for objects with the 
3.4~\micron\ feature.  Therefore, we adopt $A_K=0.2\pm0.2$ for those 
sources, which spans the extinctions for most cluster members.
In one histogram in Figure~\ref{fig:histo}, we have used photometry
in the $H$ band for the previously known members and photometry in
F162M for the new members from L24 and this work. We have measured a
median color of $H-m_{162}=0.15$ for the nonsaturated members in the
NIRCam images, which we have used to convert $m_{162}$ to $H$.
For the second histogram, we have plotted photometry in F444W when
available and otherwise have used data in the 4.5~\micron\ band from Spitzer.
We find no systematic offset between these bands for the known members.
Some members have disk emission in those bands, which will move them
to brighter magnitude bins relative to their photospheric emission. 
In Figure~\ref{fig:histo}, we have included histograms for the 19 viable 
NIRCam candidates that lack spectroscopy (Section~\ref{sec:viable}).

One notable feature of each histogram that combines the IMF
sample and the remaining NIRCam candidates is a bump near $M_{H/162}$=10--11
and $M_{4.5/444}$=8--9. This bump includes the objects that exhibit
weak-to-moderate 3.4~\micron\ absorption and may be equivalent
to the $J$-band bump that has been observed at the L/T transition among
field dwarfs \citep{dah02,tin03,vrb04,dup12}. Across that transition,
the near-IR colors of field dwarfs become bluer, which maintains a roughly
constant $J$-band magnitude. Similarly, the near-IR spectra of the
objects in IC 348 become bluer with the onset of the 3.4~\micron\ feature,
as discussed in Section~\ref{sec:class}. Unlike with field dwarfs, the
bump in IC 348 also appears in a band at longer wavelengths, F444W.

\subsubsection{Luminosity and Mass Estimates}

In \citet{luh25}, we estimated an IMF for IC 348 based on
the census of members prior to this study. For our new IMF sample,
we adopt the same mass estimates, which were derived from 
(1) positions in the Hertzsprung-Russell diagram
and evolutionary models \citep[$<$K0,][]{luh03},
(2) spectral types and a fit to dynamical mass as a function of
spectral type for young stars (K0--M7), and
(3) $M_K$ and evolutionary models \citep[][$>$M7]{bar15,cha23}.
For the new members found in L24 and this
work, we have derived masses from estimates of bolometric luminosities
in conjunction with evolutionary models.
As done in L24, we have estimated luminosities by flux calibrating
the NIRSpec data (0.6--5.3~\micron) with the NIRCam photometry, 
extrapolating to shorter and longer wavelengths with model spectra
\citep{tre15,tre17,pet23}, integrating the resulting spectra, and
assuming a distance of 313 pc. This was done for both no extinction and
an extinction correction of $A_K=0.4$, where the latter was implemented
with the extinction law from \citet{sch16av}.
For the two sources with large IR excesses in their spectra, LRL 2296 and 
LRL 11040, we estimated their photospheric fluxes at 3.5--5~\micron\ using 
other NIRSpec targets with similar 1--3~\micron\ spectra.
The luminosity estimates are included in Table~\ref{tab:spec}.
We have derived masses by combining those luminosities with the
values predicted by \citet{cha23} for an age of 5 Myr (L24).
The faintest new members have mass estimates of $\sim$2~$M_{\rm Jup}$.
The evolutionary models also imply temperatures of $\sim$900 K
for those objects.
For the viable candidates that lack spectroscopy, we have estimated
luminosities from $m_{444}$ using the correlation between that band and 
luminosity among the new members, and we have converted those luminosities
to masses using the evolutionary models. The two faintest candidates
have mass estimates of $\sim$1~$M_{\rm Jup}$.
We note that mass estimates produced by evolutionary models for young
brown dwarfs below $\sim$0.05~$M_\odot$ lack observational tests from
dynamical masses, so they could have significant systematic errors.
Nevertheless, even very large errors in the masses ($\sim50$\%) would not 
qualitatively affect our results (i.e., the detection of objects with masses
of a few $M_{\rm Jup}$).

\subsubsection{Characteristics of Substellar IMF}

\begin{figure}
\epsscale{1.2}
\plotone{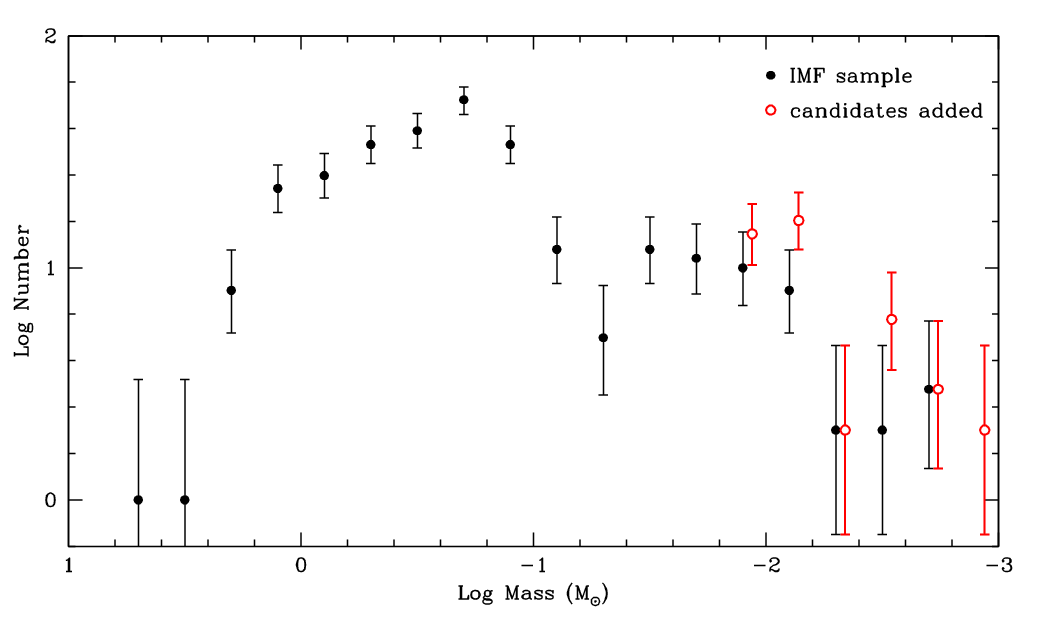}
\caption{IMF for an extinction-limited sample of members of IC 348 within
the NIRCam field (points) and the mass function after including the
remaining viable candidates from NIRCam that lack spectroscopy (open circles).
The latter points are shifted slightly in mass for clarity.}
\label{fig:imf}
\end{figure}

In Figure~\ref{fig:imf}, we present the IMF for our extinction-limited
sample of members of IC 348 in the NIRCam field, which extends
from 5 $M_\odot$ to 2~$M_{\rm Jup}$. We have included a second
histogram in which the candidates lacking spectroscopy have been added.
The mass functions are plotted with logarithmic mass bins, which results 
in a Salpeter slope of 1.35. The statistical errors are from \citet{geh86}.
The members of IC 348 with mass estimates near $\sim$2~$M_{\rm Jup}$ 
are the least massive known brown dwarfs that have spectral
classifications, providing a significant improvement in the constraint
on the minimum mass of the IMF relative to the object at 3--4~$M_{\rm Jup}$
found in L24. The two faintest candidates that lack spectroscopy suggest
that the IMF could extend to even lower masses.

The substellar IMF in Figure~\ref{fig:imf} 
exhibits a dip in the bin centered at log(mass)$=-2.3$ (5~$M_{\rm Jup}$).
When the photometric candidates are included, a modest peak also
appears in the next higher mass bin, which coincides with the bump in
the histograms of absolute magnitudes in Figure~\ref{fig:histo} and the
objects with weak-to-moderate 3.4~\micron\ bands.
This structure in the mass function may reflect errors in the mass
estimates that originate in the atmospheric and/or evolutionary models. 
Errors of this kind may not be surprising given that the
models do not predict the various spectral changes that occur
after the onset of the 3.4~\micron\ feature.
If we average across the bins at log(mass)$=-2.1$ and $-$2.3 and include
the photometric candidates, the data in Figure~\ref{fig:imf} would be
consistent with a mass function that declines very slowly down to 
log(mass)$=-2.5$ (3~$M_{\rm Jup}$), and drops somewhat faster at lower masses.
The constraints on the bottom of the IMF in IC 348 would be improved by
spectroscopy of the remaining NIRCam candidates and successful modeling
of the hydrocarbon-bearing objects.
The number ratio of stars to brown dwarfs in the IMF sample is 227/54.
If all of the 19 viable candidates lacking spectra are members, the
ratio would become 227/73.

\subsubsection{Comparison to Other Young Clusters}

In addition to IC 348, the Perseus cloud contains a second relatively
rich cluster, NGC 1333, which has $\sim$200 known members
and is younger and more heavily obscured than IC 348.
\citet{lan24} have used JWST to perform slitless spectroscopy down to
$K\sim20.5$ in an area of NGC 1333 that encompasses $\sim$50 known members,
identifying six brown dwarf candidates with $K=16.4$--19.2 and mass estimates
of 5--15~$M_{\rm Jup}$. That study concluded that it had detected the least 
massive brown dwarfs within the survey field (i.e., the minimum mass of the
IMF). However, the much deeper observations of IC 348 have uncovered
members down to masses of $\sim2$~$M_{\rm Jup}$ and two candidates near 
the mass of Jupiter. It seems likely that a deeper survey of NGC 1333 
would detect members at comparable masses.

NGC 2024 is a heavily embedded cluster in the Orion-B molecular cloud
\citep{mey08,rob24} that has been the target of a survey for brown dwarfs 
with JWST. \citet{def25} obtained very deep NIRCam images for a single 
pointing in the center of the cluster, covering two neighboring fields with 
sizes of $2\farcm2\times2\farcm2$ (the two modules of NIRCam).
That study reported a photometric sensitivity to unreddened members with
masses of 0.5~$M_{\rm Jup}$ and the detection of 48 candidates with
mass estimates as low as 3~$M_{\rm Jup}$, leading them to conclude that
they had measured the minimum mass of the IMF. However, none of the candidates
had the spectroscopy needed for confirmation of the low temperatures and
youth. In addition, coordinates and photometry were not provided
for the brown dwarf candidates, so it is not possible for 
any later study to assess their validity (with photometry or 
spectroscopy) or reproduce the mass function from that study.
In the Appendix, we perform an independent analysis of those NIRCam 
images to identify brown dwarf candidates.

\subsection{Circumstellar Disks}

Two of the new members of IC 348, LRL 2296 and LRL 11040, are significantly 
redder at $>$3.5~\micron\ in the NIRSpec data than other
cluster members at similar magnitudes, indicating the presence of excess
emission from circumstellar disks. Both objects also exhibit
absorption features that may arise from circumstellar ices 
(H$_2$O, CO$_2$, OCN$^-$, CO). The IR excesses are apparent
in Figure~\ref{fig:cmd}, where the two brown dwarfs have the reddest 
colors in $m_{162}-m_{444}$ and $m_{360}-m_{444}$ among the known members.
Given our mass estimates of $\sim2$ and 10~$M_{\rm Jup}$, the fainter one
is the least massive known brown dwarf that has evidence of a disk.
Two of the NIRCam candidates that lack spectra, LRL 11052 and LRL 11059, also
have similar colors that are suggestive of IR excesses.
Circumstellar disks often do not produce noticeable excess emission within
the wavelength range of the NIRSpec and NIRCam data, so these disk detections
provide only a lower limit for the fraction of low-mass brown dwarfs
in IC 348 that harbor disks. A thorough census of disks among these objects
would require observations at longer IR wavelengths. Data at longer
wavelengths are available in IC 348 from the Spitzer Space Telescope 
\citep{lad06,mue07}, but they are not sufficiently sensitive to detect 
members with masses of $\lesssim10$~$M_{\rm Jup}$. 

\citet{seo25} recently used a set of archival Spitzer images at 3.6 and
4.5~\micron\ to search for IR excesses among the previously
known members of IC 348 that are M9 or later. Those images had
a total exposure time of 7.5 min for a given position and filter, which
was described as ``ultradeep". All of the objects in question had been 
previously checked for IR excesses, most of them using deeper images
from Spitzer that had total exposure times of 26 min 
\citep{luh05frac,lad06,mue07}. Most of the Spitzer disk classifications
from \citet{seo25} and the earlier studies are in agreement. 
We note the few classifications that differ, and we utilize the deeper
data from NIRCam to reassess the presence of IR excesses near 4.5~\micron.
An excess was detected for LRL 5231 by \citet{seo25} but not L16.
In the NIRCam images, this object has $m_{360}-m_{444}=0.29$.
Among other members near its spectral type (M9--L0), the bluest sources,
which likely have photospheric colors, have $m_{360}-m_{444}\sim0.15$.
Thus, LRL 5231 could have a small color excess from a disk.
Meanwhile, the disk classifications for LRL 6005 and LRL 30057 also
differed between L16 and \citet{seo25}.
We measure $m_{360}-m_{444}=0.34$ for both objects.
Since both sources have uncertain spectral types (M9--L3) and 
the photospheric colors in $m_{360}-m_{444}$ become rapidly redder between
types of late M and L (Figure~\ref{fig:cmd}), it is unclear whether they have
color excesses.

We note that two of the previously known brown dwarfs
in IC 348, LRL 1843 and LRL 30003, appear as point sources that are surrounded
by extended scattered light in the NIRCam images. Both objects are located
near the protostars in the southwest corner of the cluster.

\subsection{Multiplicity}

Among the candidate and previously known brown dwarfs in IC 348 that are not 
saturated, NIRCam has resolved two close pairs that
are likely to be binary systems (Figure~\ref{fig:bin}). 
The Hubble Space Telescope has imaged some of the saturated brown dwarfs
at higher masses, none of which were resolved as binaries \citep{luh05}.
A low fraction of wide binary brown dwarfs in IC 348 is consistent 
with previous multiplicity surveys in other star-forming regions and the
solar neighborhood \citep{bur07,tod14,opi16,fon23}.

The new binary systems found with NIRCam have angular separations 
of $0\farcs23$ and $0\farcs27$, which correspond to 72 and 85 AU,
respectively, at the distance of IC 348.
Our mass estimates are $\sim$6/12~$M_{\rm Jup}$ for LRL 11043/LRL 11044
and $\sim8$/18~$M_{\rm Jup}$ for LRL 11056/LRL 1546.
These systems are rare examples of young binary brown dwarfs that
have separations of $>$10~AU and secondary masses of $<$10~$M_{\rm Jup}$.
Two other binaries of this kind are 2MASSW J1207334$-$393254 in the
TW Hya association \citep{cha04} and 2MASS J04414489+2301513 in
the Taurus star-forming region \citep{tod10}.

\section{Conclusions}

During Cycle 1 of JWST, we obtained images of a $6\farcm0\times4\farcm2$ field
in the center of IC 348 with NIRCam and we performed spectroscopy on the 
resulting brown dwarf candidates with NIRSpec (L24). Three candidates were 
confirmed to be substellar members of the cluster (3--8~$M_{\rm Jup}$). 
Two of the new members exhibited strong absorption in the 
3.4~\micron\ fundamental band and the near-IR overtone and combination
bands of an unidentified aliphatic hydrocarbon,
which were not predicted by atmospheric models and were
not previously detected in atmospheres outside of the solar system.
In Cycle 3, we have performed a deeper survey for brown dwarfs across a larger
field in IC 348 ($16\arcmin\times20\arcmin$) to test scenarios for
the origin of that hydrocarbon and to better constrain the substellar 
mass function and its minimum mass. Our results are summarized as follows:

\begin{enumerate}

\item
We have identified 39 brown dwarf candidates in IC 348 using
color-color and color-magnitude diagrams constructed from
NIRCam photometry in F162M, F182M, F360M, and F444W.
The candidates extend down to $m_{162}\sim24$, which should correspond
to masses of $\sim1$~$M_{\rm Jup}$ according to evolutionary models.

\item
The new NIRCam images have provided a second epoch of astrometry for
the $6\farcm0\times4\farcm2$ field that was observed in Cycle 1,
enabling the measurement of proper motions. The six nonsaturated sources
in the Cycle 1 field that we previously adopted as members (including the three 
new brown dwarfs found in L24) have proper motions that are tightly clustered
near the Gaia motion for IC 348, which serves as additional evidence
of their membership.

\item
We have used NIRSpec to obtain low-resolution 1--5~\micron\ spectra for 15
brown dwarf candidates and eight filler targets, where the latter
included three known members. Based on our analysis of the spectra, the 
candidates consist of nine new members of IC 348, four field T dwarfs that 
are in the background of the cluster (500--800 pc), and two active galaxies.

\item
The new NIRSpec observations have detected the 3.4~\micron\ hydrocarbon 
feature in eight of the nine new members and one previously known member.
A total of 11 members of IC 348 now have detections of it.
Among these objects, the strength of the feature is correlated with 
apparent magnitude, which indicates that the hydrocarbon is a natural 
constituent of the coolest brown dwarfs in star-forming regions.

\item
The 11 objects that have detections of the 3.4~\micron\ band exhibit
the following trends:
(1) when the 3.4~\micron\ feature first appears,
TiO and VO may begin to reemerge while the H$_2$O bands and near-IR spectral
slope remain unchanged relative to normal L dwarfs;
(2) as the 3.4~\micron\ absorption increases, TiO and VO continue
to strengthen, the H$_2$O bands weaken, and the near-IR slopes become bluer;
(3) when the 3.4~\micron\ band is strongest, the H$_2$O bands return to the 
depths of L dwarfs and the near-IR slopes may also return somewhat to redder
slopes. Some aspects of these trends are reminiscent of the L/T transition
among field dwarfs.

\item
Because of the strengths of the hydrocarbon features and the other
large spectral changes that accompany their onset, we propose 
the definition of a new spectral class (``H" for ``hydrocarbon")
that is based on the presence of the 3.4~\micron\ feature.

\item
We have used the NIRSpec data to estimate bolometric luminosities for the
new members of IC 348. Those luminosities have been converted to masses 
using evolutionary models \citep{cha23}. With mass estimates of 
$\sim$2~$M_{\rm Jup}$, the faintest new members are the least massive
known brown dwarfs, providing an improved constraint on 
the minimum mass of the IMF.

\item
Based on the results of our NIRSpec observations, five of the 24
remaining NIRCam candidates that lack spectroscopy are likely 
to be background T dwarfs or galaxies while 19 candidates have colors
that are similar to the confirmed members. Two of the latter 
candidates could have masses near the mass of Jupiter if they are members.

\item
We have estimated the IMF for an extinction-limited sample of known
members of IC 348 that are located within the NIRCam field.
The IMF is constructed with logarithmic mass bins, which
results in a Salpeter slope of 1.35.
Compared to previous samples of young stars used for IMF measurements,
the sample in IC 348 has a unique combination of relatively large size 
(281 objects), spectral classifications for all members, high level of
completeness for the selected range of extinctions, and a mass range that
spans from 5~$M_\odot$ to $\sim$2~$M_{\rm Jup}$.
The substellar IMF for our sample exhibits structure that
coincides with objects that have weak-to-moderate 3.4~\micron\ absorption,
which may reflect errors in the mass estimates that originate in the
atmospheric and evolutionary models. If we ignore that structure
and include the viable photometric candidates, the substellar IMF
appears to decline very slowly down to 3~$M_{\rm Jup}$, and drop
somewhat faster at lower masses.

\item
The NIRSpec data for two new members contain large excess emission at
$>$3.5~\micron\ from circumstellar disks. One of these objects has a 
mass estimate of $\sim2$~$M_{\rm Jup}$, making it the least massive 
known brown dwarf with evidence of a disk.

\item
Two of the new brown dwarfs have a separation of 
$0\farcs23$ (72 AU), and thus are likely to be a binary system.
In addition, one of the candidates that lacks spectroscopy has a
separation of $0\farcs27$ (85 AU) from a previously known brown dwarf.
Our mass estimates are $\sim$6/12~$M_{\rm Jup}$ for the first pair
and $\sim8$/18~$M_{\rm Jup}$ for the second pair.
These systems are rare examples of young binary brown dwarfs that have
wide separations ($>$10 AU) and secondary masses below 10~$M_{\rm Jup}$.

\end{enumerate}

\begin{acknowledgments}

We thank Donald Schneider for measuring the redshifts of the galaxies in
the NIRSpec sample.  This work is based on observations made with the 
NASA/ESA/CSA James Webb Space Telescope. The data were obtained from 
MAST at the Space Telescope Science Institute, which is operated by the 
Association of Universities for Research in Astronomy, Inc., under NASA 
contract NAS 5-03127. The JWST observations are associated with program 4866.
Support for program 4866 was provided by NASA through a grant from the Space 
Telescope Science Institute.
This work made use of ESA Datalabs (\url{https://datalabs.esa.int}), which is
an initiative by ESA's Data Science and Archives Division in the Science and
Operations Department, Directorate of Science. This work used data from the ESA
mission Gaia (\url{https://www.cosmos.esa.int/gaia}), processed by
the Gaia Data Processing and Analysis Consortium (DPAC,
\url{https://www.cosmos.esa.int/web/gaia/dpac/consortium}). Funding
for the DPAC has been provided by national institutions, in particular
the institutions participating in the Gaia Multilateral Agreement.
The Center for Exoplanets and Habitable Worlds is supported by the
Pennsylvania State University, the Eberly College of Science, and the
Pennsylvania Space Grant Consortium.

\end{acknowledgments}

\appendix

\section{Identification of Brown Dwarf Candidates in NGC 2024}
\label{sec:app}

\citet{def25} used NIRCam on JWST to obtain very deep images in the center 
of NGC 2024 through program 1190 (PI: M. Meyer) on 2023 March 1 (UT).  
The observations were performed at a single pointing using subpixel dithers.
Images were taken with eight filters that consisted of F070W, F115W, F140M, 
F182M, F356W, F360M, F430M, and F444W. The total exposure time in each
filter was 2 hr.

We retrieved the {\tt uncal} files for program 1190 from MAST:
\dataset[doi:10.17909/6t3v-px45]{http://doi.org/10.17909/6t3v-px45}.
Those data were reduced with the JWST Science Calibration pipeline.
We identified sources in the reduced images and measured aperture
photometry for them with the methods applied to our data in IC 348
(Section~\ref{sec:nircamred}).

\begin{figure}
\epsscale{0.5}
\plotone{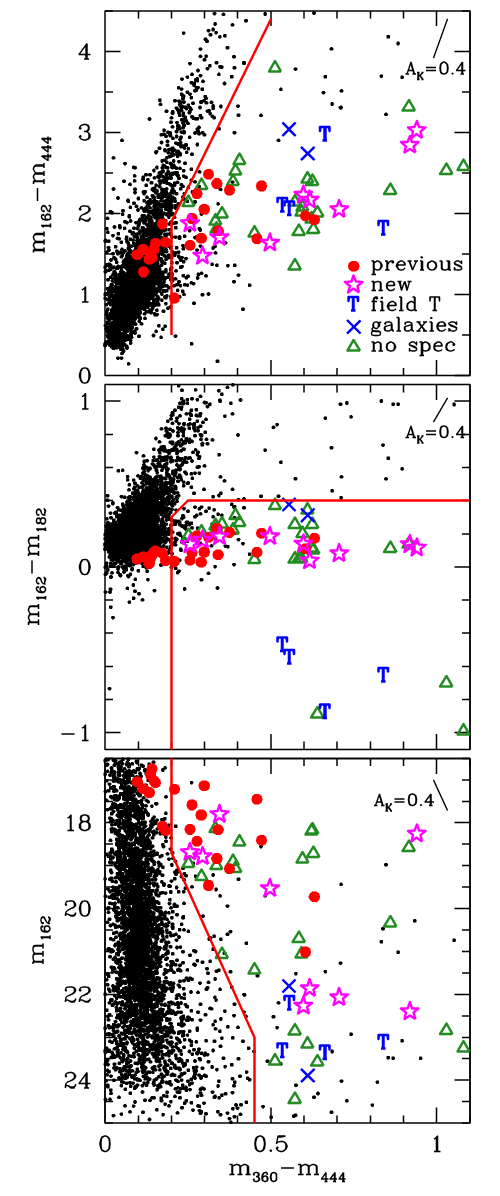}
\caption{Color-color and color-magnitude diagrams for point sources in NIRCam 
images of IC 348 from Cycle 3. We have marked the previously known 
members that are not saturated and the candidate members identified with these
diagrams, some of which have been classified as new members, field 
T dwarfs, or galaxies with NIRSpec (Table~\ref{tab:spec}).
The criteria used for selecting candidates are indicated (solid lines).}
\label{fig:cmd}
\end{figure}

\begin{figure}
\epsscale{0.6}
\plotone{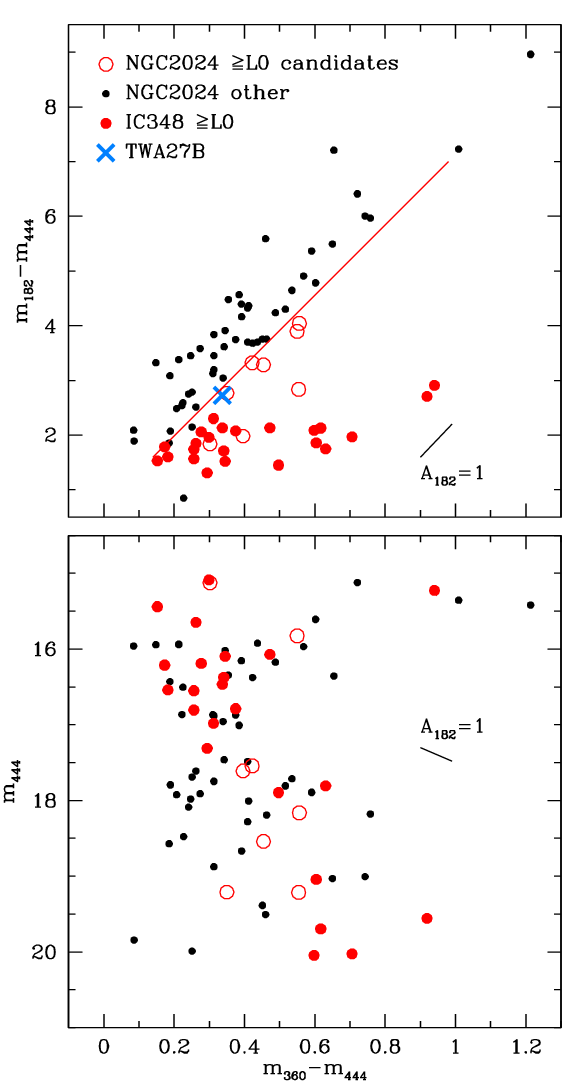}
\caption{
Color-color and color-magnitude diagrams for sources in JWST/NIRCam
images of NGC 2024 (black points and red circles), members of IC 348 with
types of $\geq$L0 (red points), and TWA 27B (cross). The latter
is shown in only the top diagram. The red line is a reddening vector
used for selecting the $\geq$L0 candidates in NGC 2024.}
\label{fig:2024}
\end{figure}

Two of the filters in the data set for NGC 2024, F140M and F182M, 
coincide with molecular absorption bands in brown dwarfs.
The combination of those bands and a filter that measures adjacent continuum,
such as F162M, can produce a color-color diagram in which brown dwarfs
have distinctive colors relative to most other astronomical sources, even in
the presence of reddening \citep[e.g.,][]{luh24o2}. Unfortunately, F162M 
was not included in program 1190, which reduces the utility of F140M and F182M.
Given the available filters, the one color-color diagram 
that is useful for the identification of brown dwarf candidates is
$m_{182}-m_{444}$ versus $m_{360}-m_{444}$, which is similar to a
diagram that was applied to IC 348 in Figure~\ref{fig:cmd}. That diagram 
is shown in Figure~\ref{fig:2024} for the point sources in NGC 2024.
We also have included a color-magnitude diagram consisting of
$m_{444}$ versus $m_{360}-m_{444}$. To help guide the selection of
brown dwarf candidates, we have marked the members of IC 348 that have
spectral types of $\geq$L0 and the young L dwarf TWA 27B \citep{luh23}.
Members of IC 348 and NGC 2024 are not perfectly comparable in
a color-magnitude diagram since they have different distances
and ages, but the differences in distances, age, and extinction roughly
cancel each other in terms of apparent magnitudes (i.e., NGC 2024 is
younger than IC 348, but it has a larger distance and higher extinctions).

In Figure~\ref{fig:2024}, we have marked a reddening vector that captures
TWA 27B and all of the $\geq$L0 members of IC 348. Two of the objects
satisfying that threshold are sufficiently underluminous in the
color-magnitude diagram relative to the members of IC 348 that we have
rejected them. The remaining eight sources below the reddening vector are
adopted as candidates for $\geq$L0 members of NGC 2024. The astrometry
and photometry for those candidates are presented in Table~\ref{tab:2024}.
Two candidates have similar colors as the IC 348 sources
while the remaining candidates have redder colors that
suggest higher extinctions. Sources 2 and 7 appear to have
the faintest extinction-corrected magnitudes, and hence are candidates
for brown dwarfs with particularly low masses.
According to APT, program 5409 has obtained NIRSpec spectra for
sources selected from the NIRCam images of NGC 2024.
It appears that four of the eight candidates that we have identified
are absent from those observations, including source 7.
Because of the high extinction in NGC 2024, some sources in the images
at F360M and F444W lack detections in bands at shorter wavelengths like F182M.
At high reddenings, we are unable to reliably distinguish brown dwarfs from 
background stars and galaxies using only the former pair of bands.

\clearpage

\begin{deluxetable}{ll} 
\tabletypesize{\scriptsize}
\tablewidth{0pt}
\tablecaption{NIRCam Data for Previously Known and Candidate Members of IC 348\label{tab:phot}}
\tablehead{
\colhead{Column Label} &
\colhead{Description}}
\startdata
LRL & LRL source name\tablenotemark{a}\\
RAdeg & Right ascension (ICRS)\\
DEdeg & Declination (ICRS)\\
m162mag & F162M NIRCam magnitude \\
e\_m162mag & Error in m162mag \\
m182mag & F182M NIRCam magnitude \\
e\_m182mag & Error in m182mag \\
m360mag & F360M NIRCam magnitude \\
e\_m360mag & Error in m360mag \\
m444mag & F444W NIRCam magnitude \\
e\_m444mag & Error in m444mag\\
memberstatus & Previously known member or candidate member
\enddata
\tablenotetext{a}{Source names are a continuation of the
designation numbers from \citet{luh98}. Sources 1, 3, and 4 from L24 are
assigned LRL numbers of 11001, 11003, and 11004, respectively.}
\tablecomments{
The table is available in its entirety in machine-readable form.}
\end{deluxetable}

\begin{deluxetable}{rlllll}
\tablecolumns{6}
\tabletypesize{\scriptsize}
\tablewidth{0pt}
\tablecaption{Sources in IC 348 Observed with NIRSpec\label{tab:spec}}
\tablehead{
\colhead{LRL\tablenotemark{a}} &
\colhead{$\alpha$ (ICRS)} &
\colhead{$\delta$ (ICRS)} &
\colhead{Membership Status} &
\colhead{Spectral Classification} &
\colhead{Log $L/L_\odot$}}
\startdata
\cutinhead{Brown Dwarf Candidates}
2121 & 56.044617 & 32.234873 & new member & H & $-3.51\pm0.17$\\
2296 & 56.021553 & 32.235174 & new member & H & $-3.29\pm0.16$\\
11024 & 55.995656 & 32.040611 & new member & H & $-4.77\pm0.12$\\
11027 & 55.980488 & 32.031945 & new member & M9--L2 & $-3.52\pm0.12$\\
11037 & 56.101440 & 32.241459 & new member & H & $-4.88\pm0.12$\\
11040 & 56.280767 & 32.257480 & new member & H & $-4.98\pm0.12$\\
11041 & 56.259664 & 32.238901 & new member & H & $-4.95\pm0.11$\\
11043 & 56.224701 & 32.230297 & new member & H & $-3.88\pm0.13$\\
11044 & 56.224665 & 32.230241 & new member & H & $-3.17\pm0.14$\\
11032 & 56.063180 & 32.215357 & background & T2  &  \nodata\\
11042 & 56.242072 & 32.262070 & background & T2  &  \nodata\\
11046 & 56.000765 & 32.165685 & background & mid-T  &  \nodata\\
11049 & 56.005416 & 32.115411 & background & T4  &  \nodata\\
11047 & 56.002391 & 32.105789 & background & galaxy (z=3.31) &  \nodata\\
11048 & 55.985803 & 32.152554 & background & galaxy (z=1.30) & \nodata \\
\cutinhead{Filler Targets}
 1546 & 56.030318 & 32.242502 & previous member & M9--L1 & $-2.78\pm0.14$\\
22705 & 56.286963 & 32.262969 & previous member & M9--L1 & $-2.90\pm0.13$\\
40013 & 56.092171 & 32.236212 & previous member & H & $-3.09\pm0.12$\\
11039 & 56.093226 & 32.244790 & background & galaxy (z=3.54) & \nodata\\
11023 & 55.993680 & 32.046580 & background & highly reddened star & \nodata\\
11025 & 55.990609 & 32.036538 & background & highly reddened star & \nodata\\
11030 & 56.016966 & 32.028483 & background & highly reddened star & \nodata\\
11031 & 56.021736 & 32.016230 & background & highly reddened star & \nodata
\enddata
\tablenotetext{a}{Source names are a continuation of the
designation numbers from \citet{luh98}.}
\end{deluxetable}

\begin{deluxetable}{rrrlllll}
\tabletypesize{\scriptsize}
\tablewidth{0pt}
\tablecaption{NIRCam Data for Candidate Members ($\geq$L0) of NGC 2024\label{tab:2024}}
\tablehead{
\colhead{ID} &
\colhead{$\alpha$ (ICRS)} &
\colhead{$\delta$ (ICRS)} &
\colhead{$m_{115}$} &
\colhead{$m_{140}$} &
\colhead{$m_{182}$} &
\colhead{$m_{360}$} &
\colhead{$m_{444}$}\\
\colhead{} &
\colhead{} &
\colhead{} &
\colhead{(mag)} &
\colhead{(mag)} &
\colhead{(mag)} &
\colhead{(mag)} &
\colhead{(mag)}}
\startdata
1 & 85.406934 & $-$1.869930 & 20.31$\pm$0.02 & 18.66$\pm$0.02 & 16.96$\pm$0.02 & 15.43$\pm$0.02 & 15.12$\pm$0.02  \\
2 & 85.412144 & $-$1.883558 & \nodata & 24.75$\pm$0.08 & 21.98$\pm$0.03 & 19.56$\pm$0.11 & 19.21$\pm$0.12  \\
3 & 85.415234 & $-$1.923661 & \nodata & \nodata & 20.87$\pm$0.02 & 17.96$\pm$0.05 & 17.54$\pm$0.04  \\
4 & 85.415437 & $-$1.896986 & \nodata & \nodata & 22.20$\pm$0.03 & 18.72$\pm$0.07 & 18.16$\pm$0.05  \\
5 & 85.416883 & $-$1.949274 & \nodata & \nodata & 21.82$\pm$0.02 & 18.99$\pm$0.06 & 18.54$\pm$0.04  \\
6 & 85.430595 & $-$1.940618 & 22.22$\pm$0.02 & 21.11$\pm$0.02 & 19.59$\pm$0.02 & 18.01$\pm$0.02 & 17.61$\pm$0.02  \\
7 & 85.431232 & $-$1.867624 & 25.35$\pm$0.04 & 24.08$\pm$0.03 & 22.05$\pm$0.02 & 19.77$\pm$0.06 & 19.21$\pm$0.03  \\
8 & 85.439093 & $-$1.928280 & 22.87$\pm$0.02 & 21.84$\pm$0.02 & 19.72$\pm$0.02 & 16.38$\pm$0.02 & 15.83$\pm$0.02  \\
\enddata
\end{deluxetable}

\clearpage

\begin{thebibliography}{}




\bibitem[Allers \& Liu(2020)]{all20}
Allers, K. N., \& Liu, M. C. 2020, \pasp, 132, 104401

\bibitem[Alves de Oliveira et al.(2018)]{alv18}
Alves de Oliveira, C., Birkmann, S. M., B{\"o}ker, T., et al. 2018, SPIE, 
10704, 107040Q


\bibitem[Alves de Oliveira et al.(2013)]{alv13}
Alves de Oliveira, C., Moraux, E., Bouvier, J., et al. 2013, \aap, 549, A123



\bibitem[Baraffe et al.(2015)]{bar15}
Baraffe, I., Hormeier, D., Allard, F., \& Chabrier, G. 2015, \aap, 577, 42


\bibitem[Basri et al.(1996)]{bas96}
Basri, G., Marcy, G. W., \& Graham, J. R. 1996, \apj, 458, 600

\bibitem[Belluci et al.(2009)]{bel09}
Bellucci, A., Sicardy, B., Drossart, P., et al. 2009, \icarus, 201, 198







\bibitem[Burgasser et al.(2006)]{bur06}
Burgasser, A. J., Geballe, T. R., Leggett, S. K., Kirkpatrick, J. D.,  
\& Golimowski, D. A. 2006, \apj, 637, 1067

\bibitem[Burgasser et al.(2002a)]{bur02a} 
Burgasser, A. J., Kirkpatrick, J. D., Brown, M. E., et al. 2002a, \apj, 564, 421


\bibitem[Burgasser et al.(2002b)]{bur02b} 
Burgasser, A. J., Marley, M. S., Ackerman, A. S., et al. 2002b, \apjl, 571, L151

\bibitem[Burgasser et al.(2007)]{bur07} 
Burgasser, A. J., Reid, I. N., Siegler, N., et al. 2007, in Protostars and 
Planets V, ed. V. B. Reipurth, D. Jewitt, \& K. Keil (Tucson, AZ: Univ. 
Arizona Press), 427






\bibitem[Chabrier et al.(2023)]{cha23}
Chabrier, G., Baraffe, I., Phillips, M., \& Debras, F. 2023, \aap, 671, A119



\bibitem[Chauvin et al.(2004)]{cha04}
Chauvin, G., Lagrange, A.-M., Dumas, C., et al. 2004, \aap, 425, L29








\bibitem[Cruz et al.(2009)]{cru09}  
Cruz, K. L., Kirkpatrick, J. D., \& Burgasser, A. J. 2009, \aj, 137, 3345


\bibitem[Cushing et al.(2011)]{cus11}
Cushing, M. C., Kirkpatrick, J. D., Gelino, C. R., et al. 2011, \apj, 743, 50



\bibitem[Dahn et al.(2002)]{dah02}
Dahn, C. C., Harris, H. C., Vrba, F. J., et al. 2002, \aj, 124, 1170




\bibitem[De Furio et al.(2025)]{def25}
De Furio, M., Meyer, M. R., Green, T., et al. 2025, \apjl, 981, L34



\bibitem[Dupuy et al.(2012)]{dup12}
Dupuy, T. J., \& Liu, M. C. 2012, \apjs, 201, 19


\bibitem[Esplin \& Luhman(2017)]{esp17}
Esplin, T. L., \& Luhman, K. L. 2017, \aj, 154, 134




\bibitem[Ferruit et al.(2022)]{fer22}
Ferruit, P., Jakobsen, P., Giardino, G., et al. 2022, \aap, 661, A81


\bibitem[Fontanive et al.(2023)]{fon23}
Fontanive, C., Bedin, L. R., De Furio, M., et al. 2023, \mnras, 526, 1783


\bibitem[Gaia Collaboration et al.(2021)]{bro21}
Gaia Collaboration, Brown, A. G. A., Vallenari, A., et al. 2021, \aap, 649, A1

\bibitem[Gaia Collaboration et al.(2016)]{gaia16b}
Gaia Collaboration, Prusti, T., de Bruijne, J. H. J., et al. 2016, \aap, 595, A1

\bibitem[Gaia Collaboration et al.(2023)]{val23}
Gaia Collaboration, Vallenari, A., Brown, A. G. A., et al. 2023, \aap, 674, A1

\bibitem[Gardner et al.(2023)]{gar23} 
Gardner, J. P., Mather, J. C., Abbott, R., et al. 2023, \pasp, 135, 068001


\bibitem[Geballe et al.(2002)]{geb02}
Geballe, T. R., Knapp, G. R., Leggett, S. K., et al. 2002, \apj, 564, 466



\bibitem[Gehrels(1986)]{geh86}
Gehrels, N. 1986, \aj, 303, 336






\bibitem[Herbig(1998)]{her98} 
Herbig, G. H. 1998, \apj, 497, 736

\bibitem[Herbst(2008)]{her08}
Herbst, W. 2008, in Handbook of Star Forming Regions, Vol. 1, ed. B. Reipurth 
(San Francisco, CA: ASP), 372

\bibitem[Hillenbrand(1997)]{hil97}
Hillenbrand, L. A. 1997, \aj, 113, 1733




\bibitem[Jakobsen et al.(2022)]{jak22}
Jakobsen, P., Ferruit, P., Alves de Oliveira, C., et al. 2022, \aap, 661, A80





\bibitem[Kirkpatrick(2005)]{kir05}
Kirkpatrick, J. D. 2005, \araa, 43, 195



\bibitem[Kirkpatrick et al.(1999)]{kir99}
Kirkpatrick, J. D., Reid, I. N., Liebert, J., et al. 1999, \apj, 519, 802



\bibitem[Lada \& Lada(1995)]{lad95}
Lada, E. A., \& Lada, C. J. 1995, \aj, 109, 1682

\bibitem[Lada et al.(2006)]{lad06}
Lada, C. J., Muench, A. A., Luhman, K. L., et al. 2006, \aj, 131, 1574

\bibitem[Lalchand et al.(2022)]{lal22}
Lalchand, B., Chen, W.-P., Biller, B. A., et al. 2022, \aj, 164, 125

\bibitem[Langeveld et al.(2024)]{lan24}
Langeveld, A. B., Scholz, A., Mu{\v{z}}i{\'c}, K., et al. 2024, \aj, 168, 179






\bibitem[Lucas et al.(2001)]{luc01}
Lucas, P. W., Roche, P. F., Allard, F., \& Hauschildt, P. H. 2001, \mnras,
326, 695

\bibitem[Luhman(1999)]{luh99}
Luhman, K. L. 1999, \apj, 525, 466



\bibitem[Luhman(2024)]{luh24o2}
Luhman, K. L. 2024, \aj, 168, 230

\bibitem[Luhman(2025)]{luh25}
Luhman, K. L. 2025, \aj, in press

\bibitem[Luhman et al.(2024)]{luh24ic}
Luhman, K. L., Alves de Oliveira, C., Baraffe, I., et al. 2024, \aj, 167, 19



\bibitem[Luhman et al.(2016)]{luh16}
Luhman, K. L., Esplin, T. E., \& Loutrel, N. P. 2016, \apj, 827, 52

\bibitem[Luhman \& Hapich(2020)]{luh20ic}
Luhman, K. L., \& Hapich, C. J. 2020, \aj, 160, 57

\bibitem[Luhman et al.(2005a)]{luh05frac}
Luhman, K. L., Lada, E. A., Hartmann, L., et al. 2005a, \apj, 631, L69

\bibitem[Luhman et al.(2005b)]{luh05flam}
Luhman, K. L., Lada, E. A., Muench, A. A., \& Elston, R. J. 2005a, \apj, 618,
810

\bibitem[Luhman et al.(1997)]{luh97}
Luhman, K. L., Liebert, J., \& Rieke, G. H. 1997, \apjl, 489, L165

\bibitem[Luhman et al.(2017)]{luh17}
Luhman, K. L., Mamajek, E. E., Shukla, S. J., \& Loutrel, N. P. 2017, \aj,
153, 46

\bibitem[Luhman et al.(2005c)]{luh05}
Luhman, K. L., McLeod, K. K., \& Goldenson, N. 2005c, \apj, 623, 1141

\bibitem[Luhman et al.(1998)]{luh98}
Luhman, K. L., Rieke, G. H., Lada, C. J., \& Lada, E. A. 1998, \apj, 508, 347

\bibitem[Luhman et al.(2003)]{luh03}
Luhman, K. L., Stauffer, J. R., Muench, A. A., et al. 2003, \apj, 593, 1093

\bibitem[Luhman et al.(2023)]{luh23}
Luhman, K. L., Tremblin, P., Birkmann, S. M., et al. 2023, \apjl, 949, L36






\bibitem[Mart{\'\i}n et al.(1997)]{mar97}
Mart{\'\i}n, E. L., Basri, G., Delfosse, X., \& Forveille, T. 1997, \aap,
327, L29

\bibitem[Matrajt et al.(2005)]{mat05}
Matrajt, G., Mu{\~n}oz Caro, G. M., Dartois, E., et al. 2005, \aap, 433, 979




\bibitem[Meyer et al.(2008)]{mey08}
Meyer, M. R., Flaherty, K., Levine, J. L., et al. 2008, in Handbook of Star
Forming Regions, Vol. 1, ed. B. Reipurth (San Francisco, CA: ASP), 662

\bibitem[Muench et al.(2007)]{mue07}
Muench, A. A., Lada, C. J., Luhman, K. L., Muzerolle, J., \& Young, E. 2007,
\aj, 134, 411




\bibitem[Nakajima et al.(1995)]{nak95}
Nakajima, T., Oppenheimer, B. R., Kulkarni, S. R., et al. 1995, \nat, 378, 463

\bibitem[Opitz et al.(2016)]{opi16}
Opitz, D., Tinney, C. G., Faherty, J., et al. 2016, \apj, 819, 17

\bibitem[Oppenheimer et al.(1995)]{opp95}
Oppenheimer, B. R., Kulkarni, S. R., Nakajima, T., \& Matthews, K. 1995,
Science, 270, 1478




\bibitem[Pendleton \& Allamandola(2002)]{pen02}
Pendleton, Y. J., \& Allamandola, L. J. 2002, \apjs, 138, 75

\bibitem[Pendleton et al.(1994)]{pen94}
Pendleton, Y. J., Sandford, S. A., Allamandola, L. J., Tielens, A. G. G. M.,
\& Sellgren, K. 1994, \apj, 437, 683



\bibitem[Petrus et al.(2023)]{pet23} 
Petrus, S., Chauvin, G., Bonnefoy, M., et al. 2023, \aap, 670, L9


\bibitem[Rebolo et al.(1996)]{reb96}
Rebolo, R.,  Mart{\'\i}n, E. L., Basri, G., et al. 1996, \apj, 469, L53

\bibitem[Rebolo et al.(1995)]{reb95}
Rebolo, R., Zapatero Osorio, M. R., \& Mart{\'\i}n, E. L. 1995, \nat, 377, 129

\bibitem[Rieke et al.(2005)]{rie05}
Rieke, M. J., Kelly, D. M., \& Horner, S. 2005, SPIE, 5904, 590401

\bibitem[Rieke et al.(2023)]{rie23}
Rieke, M. J., Kelly, D. M., Misselt, K., et al. 2023, \pasp, 135, 028001

\bibitem[Robberto et al.(2024)]{rob24}
Robberto, M., Gennaro, M., Da Rio, N., et al. 2024, \apj, 960, 49

\bibitem[Robin et al.(2003)]{rob03}
Robin, A. C., Reyl{\'e}, C. Derri{\`e}re, S., \& Picaud, S. 2003, \aap, 409,
523





\bibitem[Sandford et al.(1991)]{san91}
Sandford, S. A., Allamandola, L. J., Tielens, A. G. G. M., et al. 1991, \apj,
371, 607

\bibitem[Schlafly et al.(2016)]{sch16av}
Schlafly, E. F., Meisner, A. M., Stutz, A. M., et al. 2016, \apj, 821, 78




\bibitem[Seo \& Scholz(2025)]{seo25}
Seo, H. H., \& Scholz, A. 2025, \mnras, 537, 2579

\bibitem[Soifer et al.(1976)]{soi76}
Soifer, B. T., Russell, R. W., \& Merrill, K. M. 1976, \apjl, 207, L83


\bibitem[Stauffer et al.(1994)]{sta94}
Stauffer, J. R., Hamilton, D., \& Probst, R. G. 1994, \aj, 108, 155



\bibitem[Tinney et al.(2003)]{tin03}
Tinney, C. G., Burgasser, A. J., \& Kirkpatrick, J. D. 2003, \aj, 126, 975

\bibitem[Todorov et al.(2010)]{tod10}
Todorov, K., Luhman, K. L., \& McLeod, K. K. 2010, \apjl, 714, L84

\bibitem[Todorov et al.(2014)]{tod14}
Todorov, K., Luhman, K. L., Konopacky, Q. M., et al. 2014, \apj, 788, 40


\bibitem[Tremblin et al.(2015)]{tre15}
Tremblin, P., Amundsen, D. S., Mourier, P., et al. 2015, \apjl, 804, L17

\bibitem[Tremblin et al.(2017)]{tre17}
Tremblin, P., Chabrier, G., Baraffe, I., et al. 2017, \apj, 850, 46


\bibitem[Vrba et al.(2004)]{vrb04}
Vrba, F. J., Henden, A. A., Luginbuhl, C. B., et al. 2004, \aj, 127, 2948

\bibitem[Wdowiak et al.(1988)]{wdo88}
Wdowiak, T. J., Flickinger G. C., \& Cronin J. R. 1988, \apjl, 328, L75







\bibitem[Zhang et al.(2025)]{zha25}
Zhang, Z., Molli\'{e}re, P., Fortney, J., \& Marley, M. S. 2025, \aj,
submitted

\end{thebibliography}
\end{document}